\documentclass[preprint,12pt]{elsarticle}
\journal{International Journal of Engineering Sciences}

\usepackage{graphics,epsfig}
\usepackage{graphicx}
\usepackage{float}
\usepackage{amssymb,amstext,amsmath}
\usepackage{mathtools}
\mathtoolsset{showonlyrefs}
\usepackage{booktabs}
\usepackage{natbib}
\usepackage[latin1]{inputenc}
\usepackage{epstopdf}

\begin{document}
\begin{frontmatter}
\title{An asymptotically exact theory of smart sandwich shells}
\author{K. C. Le$^a$\footnote{Corresponding author: phone: +49 234 32-26033, email: chau.le@rub.de.}, J. H. Yi$^b$}
\address{$^a$Lehrstuhl f\"{u}r Mechanik - Materialtheorie, Ruhr-Universit\"{a}t Bochum,\\D-44780 Bochum, Germany\\
$^b$Institute of Complex Systems, Forschungszentrum J\"ulich, D-52425 J\"ulich, Germany}

\begin{abstract} 
An asymptotically exact two-dimensional theory of elastic-piezoceramic sandwich shells is derived by the variational-asymptotic method. The error estimation of the constructed theory is given in the energetic norm. As an application, analytical solution to the problem of forced vibration of a circular elastic plate partially covered by two piezoceramic patches with thickness polarization excited by a harmonic voltage is found. 
\end{abstract}

\begin{keyword}
elastic, piezoelectric, sandwich, shell, variational-asymptotic method.
\end{keyword}
\end{frontmatter}

\section{Introduction}

In the past thirty years piezoelectric materials has been widely used as sensors or actuators for the active vibration control of various smart structures like beams \citep{Bailey1985,Crawley1987}, plates \citep{Crawley1991}, and shells \citep{Tzou1989}. In typical situations one or two piezoelectric patches are attached to one or both sides of the elastic  structure to be controlled \citep{Preumont2011}. If these patches are excited by an oscillating voltage, they contract or elongate and thereby exert, depending on the configuration of patches and electrodes, membrane force or bending moment on the structure. In the area of bonded patches the structure changes from homogeneous through the thickness to a sandwich structure with two or three layers: one elastic layer and one or two piezoelectric patches perfectly bonded to it. Modern piezoelectric devices, such as harvesters \citep{Anton2007}, require more complicated piezo patch arrangements. Thus, shape and topology optimization of piezo patches integrated into elastic structures becomes a challenging open field of research, in particular for thin smart structures (see, eg., \citep{Wein2009} and the references therein).

Due to the above mentioned complicated laminate structure, the problems of equilibrium and vibration of smart sandwich plates and shells admit exact analytical solutions of the three-dimensional theory of piezoelectricity only in a few exceptional cases (see, e.g., \citep{Pan2001}). By this reason different approaches have been developed depending on the type of the structures. If smart sandwich plates and shells are thick, no accurate two-dimensional  theory can be constructed, so only the numerical methods applied to three-dimensional theory of piezoelectricity make sense \citep{Allik1970,He2002}. However, if smart plates and  shells are thin, the reduction from the three- to two-dimensional theory is possible and different approximations can be constructed. Up to now two main approaches have been developed: (i) the variational approach based on Hamilton's variational principle and on some ad-hoc assumptions generalizing Kirchhoff-Love's hypothesis to smart sandwich plates and shells \citep{Lee1990,Tzou1993,Tzou1994,Zhang1999,Benjeddou2002}\footnote{The literature on this topics is huge due to the variety of the 2-D smart sandwich shell and plate theories: single-layer, multi-layer, refined theories including rotary inertias and transverse shears et cetera. It is therefore impossible to cite all references. For the overview the reader may consult \citep{Saravanos1999,Tzou2012,Yu2012} and the references therein.}, (ii) the asymptotic approach based on the analysis of the three-dimensional equations of piezoelectricity, mainly for the plates \citep{Maugin1990,Cheng2000a,Cheng2000b,Leugering2012}. The disadvantage of the variational approach is the necessity of having an Ansatz for the displacements and electric field that is difficult to be justified, while simplicity and brevity are its advantages. The asymptotic method needs no a priori assumptions; however, the direct asymptotic analysis of the 3-D differential equations of piezoelectricity is very cumbersome. The synthesis of these two approaches, called the variational-asymptotic method, first proposed by \citet{Berdichevsky1979} and developed further in \citep{Le1999}, seems to avoid the disadvantages of both approaches described above and proved to be quite effective in constructing approximate equations for thin-walled structures. Note that this method has been applied, among others, to derive the 2-D theory of homogeneous piezoelectric shells in \citep{Le1986a} and the 2-D theory of purely elastic sandwich plates and shells in \citep{Berdichevsky2010a,Berdichevsky2010b}. Note also the closely related method of  gamma convergence used in homogenization \citep{Braides2002} and plate theories \citep{Friesecke2006}.

The aim of this paper is to construct the rigorous first order approximate two-dimensional smart  sandwich shell theory by the variational-asymptotic method. We consider the sandwich shell with one elastic layer in the middle and two piezoceramic patches symmetrically bonded to it. The dimension reduction is based on the asymptotic analysis of the action functional containing small parameters that enables one to find the distribution of the displacements and electric field from the solution of the so-called thickness problem. Using the generalized Prager-Synge identity for the inhomogeneous piezoelectric body, we provide also the error estimation of the constructed theory in the energetic norm. We apply this theory to the problem of forced vibration of a circular plate partially covered by two circular rings of piezoceramic layers with thickness polarization excited by a harmonic voltage and find the exact analytical solution to this problem.

The paper is organized as follows. After this short introduction the variational formulation of the problem is given in Section 2. Sections 3 and 4 are devoted to the asymptotic analysis of the action functional. In Section 5 the two-dimensional theory of piezoelectric sandwich shells is obtained. In Section 6 we provide the error estimation of the constructed theory. Section 7 presents the exact analytical solution to the forced vibration of a circular sandwich plate. Finally, Section 8 concludes the  paper.
 
\section{Variational formulation of the problem}
Let $\Omega $ be a two-dimensional smooth surface bounded by a smooth closed curve $\partial \Omega $. At each point of the surface $\Omega $ (called the middle surface of the shell) a  segment of length $h$ in the direction perpendicular to the surface is drawn so that its centre lies on the surface. If the length $h$ is sufficiently small, the segments do not intersect each other and fill the domain $\mathcal{V}$ occupied by a sandwich shell in its undeformed state. 
\begin{figure}[htb]
	\centering
	\includegraphics[width=10cm]{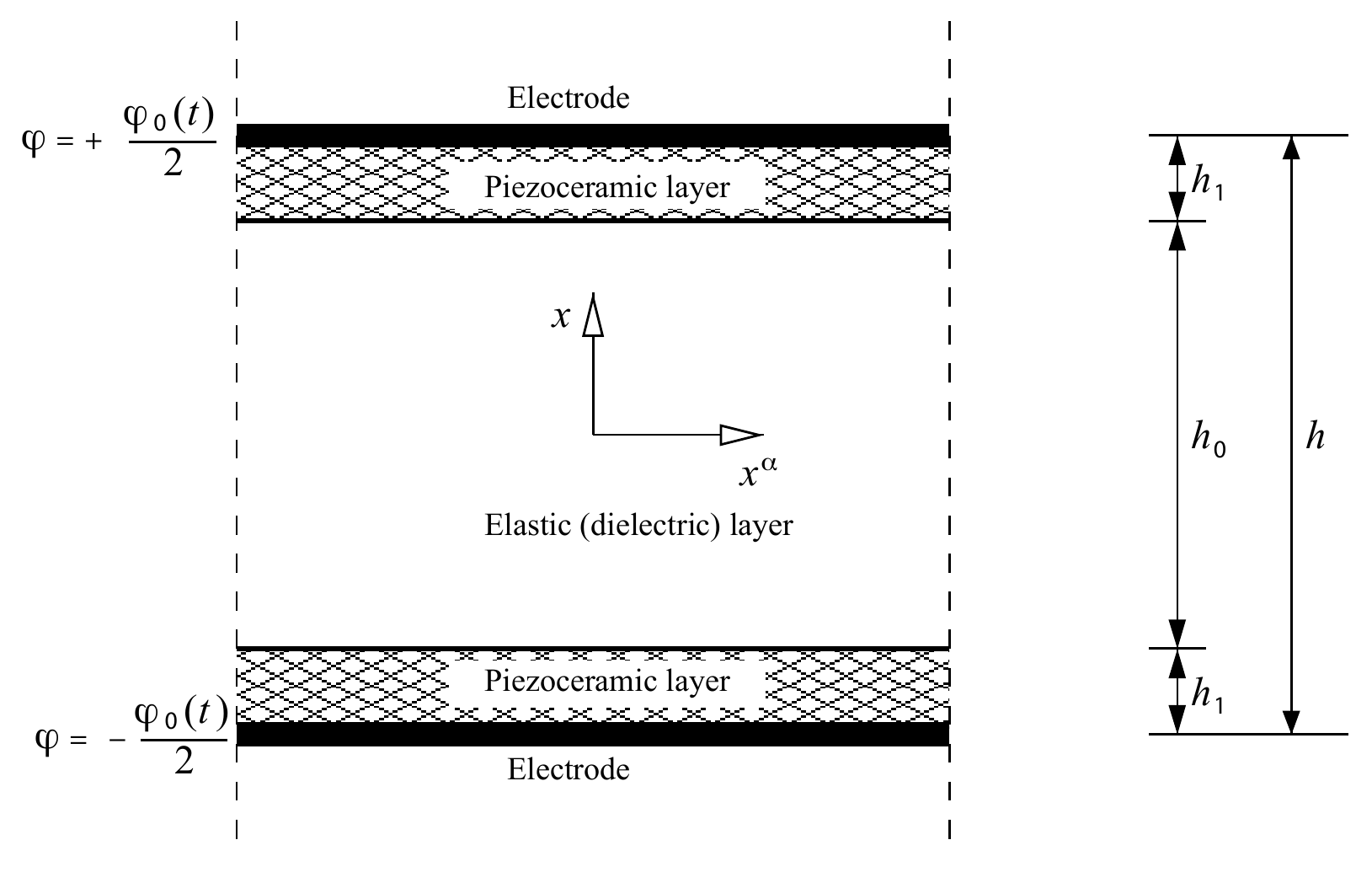}
	\caption{A side view of the sandwich shell}
	\label{fig:1}
\end{figure}
We analyze the forced vibration of the sandwich shell consisting of three layers shown schematically in Fig.~\ref{fig:1}. The middle layer is made of an elastic (dielectric) transversely isotropic and homogeneous material that does not exhibit piezo-effects. The upper and lower layers are made of the same transversely isotropic homogeneous piezoceramic material with the thickness polarization perfectly attached to the middle one. These layers can be used as sensor or actuator for the purpose of active control of shell vibration. The faces of piezoelectric layers are covered by the electrodes whose thickness is neglected. Thus, the undeformed shell  occupies the domain specified by the equation
\begin{displaymath}
z^i(x^\alpha ,x)=r^i(x^\alpha )+xn^i(x^\alpha ),
\end{displaymath}
where $z^i=r^i(x^\alpha )$ is the equation of the middle surface $\Omega $, and $n^i(x^\alpha )$ are the cartesian components of the normal vector ${\bf n}$ to this surface. We shall use Latin indices running from 1 to 3 to refer to the spatial co-ordinates and the Greek indices running from 1 to 2 to refer to the surface co-ordinates $x^1$ and $x^2$. The curvilinear co-ordinates $x^\alpha $ take values in a domain of $\mathbb{R}^2$, while $x\in [-h/2,h/2]$. Let the upper and lower faces of the piezoceramic layers covered by the electrodes (at $x=\pm h/2$) be denoted by $\Omega _\pm$. On these surfaces the electric potential is prescribed
\begin{equation}
\label{2.1}
\varphi = \pm \varphi _0(t)/2 \quad \text{on $\Omega _\pm$}.
\end{equation}

Hamilton's variational principle of piezoelectricity (see, e.g., \citep{Le1986a,Le1999}) states that the true displacement $\check{\mathbf{w}}(\mathbf{x},t)$ and electric potential $\check{\varphi }(\mathbf{x},t)$ of an inhomogeneous piezoelectric body change in space and time in such a way that the action functional
\begin{equation}
\label{2.2}
I[\mathbf{w}(\mathbf{x},t),\varphi (\mathbf{x},t)]=\int_{t_0}^{t_1}\int_{\mathcal{V}}[T(\mathbf{x},\dot{\mathbf{w}})-W(\mathbf{x},\boldsymbol{\varepsilon} ,\mathbf{E})]\, dv \, dt
\end{equation} 
becomes stationary among all continuously differentiable functions $\mathbf{w}(\mathbf{x},t)$ and $\varphi (\mathbf{x},t)$ satisfying the initial and end conditions
\begin{displaymath}
\mathbf{w}(\mathbf{x},t_0)=\mathbf{w}_0(\mathbf{x}), \quad \mathbf{w}(\mathbf{x},t_1)=\mathbf{w}_1(\mathbf{x})
\end{displaymath}
as well as constraint \eqref{2.1}. The integrand in the action functional \eqref{2.2} is called Lagrangian, while $dv$ is the volume element and the dot over quantities denotes the partial time derivative. In the Lagrangian $T(\mathbf{x},\dot{\mathbf{w}})$ describes the kinetic energy density given by\footnote{As we shall be concerned with mechanical vibrations of non-conducting piezoelectric bodies at frequencies far below optical frequencies, the coupling between the electric and magnetic fields and the dependence of the kinetic energy on $\dot{\varphi }$ can be neglected.}
\begin{equation}
\label{2.3}
T(\mathbf{x},\dot{\mathbf{w}})=\frac{1}{2}\rho (\mathbf{x}) \dot{\mathbf{w}} \cdot \dot{\mathbf{w}},
\end{equation}
with $\rho (\mathbf{x})$ being the mass density. Function $W(\mathbf{x},\boldsymbol{\varepsilon} ,\mathbf{E})$, called electric enthalpy density, reads 
\begin{equation}
\label{2.4}
W(\mathbf{x},\boldsymbol{\varepsilon} ,\mathbf{E})=\frac{1}{2}\boldsymbol{\varepsilon} \mathbf{:}\mathbf{c}(\mathbf{x})\mathbf{:}\boldsymbol{\varepsilon}-\mathbf{E}\cdot \mathbf{e}(\mathbf{x})\mathbf{:}\boldsymbol{\varepsilon}-\frac{1}{2}\mathbf{E}\cdot \boldsymbol{\epsilon}(\mathbf{x})\cdot \mathbf{E},
\end{equation}
where $\boldsymbol{\varepsilon} $ is the strain tensor
\begin{equation}
\label{2.4a}
\boldsymbol{\varepsilon }=\frac{1}{2}(\nabla \mathbf{w}+(\nabla \mathbf{w})^T),
\end{equation}
while $\mathbf{E}$ the electric field
\begin{equation}
\label{2.4b}
\mathbf{E}=-\nabla \varphi .
\end{equation}
Applying the standard calculus of variation one easily shows that the stationarity condition $\delta I=0$ implies the equations of motion of piezoelectric body (including the equation of electrostatics)
\begin{equation}
\label{2.5}
\rho (\mathbf{x})\ddot{\mathbf{w}}=\text{div}\boldsymbol{\boldsymbol{\sigma }}, \quad \text{div}\mathbf{D}=0,
\end{equation}
where the stress tensor $\boldsymbol{\boldsymbol{\sigma }}$ and the electric induction field $\mathbf{D}$ are given by
\begin{equation}
\label{2.6}
\begin{split}
\boldsymbol{\boldsymbol{\sigma }}=\frac{\partial W}{\partial \boldsymbol{\varepsilon}}=\mathbf{c}(\mathbf{x})\mathbf{:}\boldsymbol{\varepsilon}-\mathbf{E}\cdot \mathbf{e}(\mathbf{x}),
\\
\mathbf{D}=-\frac{\partial W}{\partial \mathbf{E}}=\mathbf{e}(\mathbf{x})\mathbf{:}\boldsymbol{\varepsilon}+\boldsymbol{\epsilon}(\mathbf{x})\cdot \mathbf{E}.
\end{split}
\end{equation}
We call $\mathbf{c}(\mathbf{x})$ the (fourth-rank) tensor of elastic stiffnesses, $\mathbf{e}(\mathbf{x})$ the (third-rank) tensor of piezoelectric constants, while $\boldsymbol{\epsilon}(\mathbf{x})$ the (second-rank) tensor of dielectric permittivities. For the elastic (dielectric) material $\mathbf{e}=0$, so it is the degenerate case of piezoelectric material. Substituting the constitutive equations \eqref{2.6} into \eqref{2.5} and making use of the kinematic equations \eqref{2.4a} and \eqref{2.4b}, we get the closed system of four governing equations for four unknown functions $\mathbf{w}(\mathbf{x},t)$ and $\varphi (\mathbf{x},t)$.

For the asymptotic analysis of the sandwich shell it is convenient to use the curvilinear coordinates $\{ x^\alpha ,x\}$ introduced above and the co- and contravariant index notation for vectors and tensors, with Einstein's summation convention being employed. In this coordinate system the action functional reads 
\begin{equation}\label{2.7}
I=\int_{t_0}^{t_1}\int_{\Omega}\int_{-h/2}^{h/2}[T(x,\dot{\mathbf{w}})-W(x,\boldsymbol{\varepsilon} ,\mathbf{E})] \kappa \, dx\, da\, dt,
\end{equation}
where $\kappa =1-2Hx+Kx^2$ (with $H$ and $K$ being the mean and Gaussian curvature of the middle surface, respectively) and $da$ denotes the area element of the middle surface. The kinetic energy becomes 
\begin{equation*}
T(x,\dot{\mathbf{w}})=\frac{1}{2}\rho (x) (a^{\alpha \beta }\dot{w}_\alpha \dot{w}_\alpha +\dot{w}^2),
\end{equation*}
where $a^{\alpha \beta }$ are the contravariant components of the surface metric tensor, while $w_\alpha $ and $w$ are the projections of the displacement vector onto the tangential and normal directions to the middle surface
\[
w_\alpha =t^i_\alpha w_i=r^i_{,\alpha }w_i,\quad w=n^iw_i.
\]
The electric enthalpy density $W$ reads
\begin{equation*}
W(x,\boldsymbol{\varepsilon} ,\mathbf{E})=\frac{1}{2}c^{abcd}(x)\varepsilon _{ab}\varepsilon _{cd} -e^{cab}(x)\varepsilon _{ab}E_c -\frac{1}{2}\epsilon ^{ab}(x)E_a E_b.
\end{equation*}
For the sandwich shells we have
\begin{align*}
\rho (x)&=\begin{cases}
   \rho   & \text{for $|x|>h_0/2$}, \\
   \bar{\rho }  & \text{for $|x|<h_0/2$},
   \end{cases}
\\
c^{abcd}(x)&=\begin{cases}
   c_E^{abcd}   & \text{for $|x|>h_0/2$}, \\
   \bar{c}^{abcd}   & \text{for $|x|<h_0/2$},
\end{cases}
\end{align*}
\begin{align*}
e^{cab}(x)&=\begin{cases}
   e^{cab}   & \text{for $|x|>h_0/2$}, \\
  0  & \text{for $|x|<h_0/2$},
\end{cases}
\\
\epsilon^{ab}(x)&=\begin{cases}
   \epsilon_S^{ab}   & \text{for $|x|>h_0/2$}, \\
   \bar{\epsilon}^{ab}   & \text{for $|x|<h_0/2$}.
\end{cases}
\end{align*}
Thus, the label $E$ in $c_E^{abcd}$ indicates elastic stiffnesses at constant electric field, while the label $S$ in $\epsilon_S^{ab}$ denotes dielectric permittivities at constant strain for the piezoelectric material. The material constants with an over-bar correspond to those of the elastic (dielectric) material. 

The problem is to replace the three-dimensional action functional \eqref{2.7} by an approximate two-dimensional action functional for a thin shell, whose functions depend only on the longitudinal co-ordinates $x^1,x^2$ and time $t$. The possibility of reduction of the three- to the two-dimensional problem is related to the smallness of the ratios between the thickness $h$ and the characteristic radius of curvature $R$ of the shell middle surface and between $h$ and the characteristic scale of change of the electroelastic state in the longitudinal directions $l$ \citep{Le1999}. We assume that
\begin{displaymath}
\frac{h}{R}\ll 1, \quad \frac{h}{l}\ll 1.
\end{displaymath}
Additionally, we assume that 
\begin{equation}
\frac{h}{ c\tau }\ll 1,
\label{2.7a}
\end{equation}
where $\tau $ is the characteristic scale of change of the functions $w_i$ and $\varphi $ in time (see \citep{Le1999}) and $c$ the minimal velocity of plane waves in the piezoelectric materials under consideration. This means that we consider in this paper only statics or low-frequency vibrations of the inhomogeneous piezoelectric shell. By using the variational-asymptotic method, the two-dimensional action functional will be constructed below in which terms of the order $h/R$ and $h/l$ are neglected as compared with unity (the first-order or ``classical'' approximation).

In order to fix the domain of the transverse co-ordinate in the passage to the limit $h\to 0$, we introduce the dimensionless co-ordinate
\begin{equation*}
\zeta =\frac{x}{h}, \quad \zeta \in [-1/2,1/2].
\end{equation*}
Now $h$ enters the action functional explicitly through the components of the strain tensor $\varepsilon _{ab}$ and the electric field $E_a$ 
\begin{align}
\varepsilon _{\alpha \beta }&=
w_{(\alpha ;\beta )}-b_{\alpha \beta }w-
h\zeta b^\lambda _{(\alpha }w_{\lambda ;\beta )}+h\zeta 
c_{\alpha \beta }w,
\notag \\
2\varepsilon _{\alpha 3}&= 
\frac{1}{h}w_{\alpha |\zeta }+w_{,\alpha }+b^\lambda _\alpha w_\lambda 
-\zeta b^\lambda _\alpha w_{\lambda |\zeta }, \quad \varepsilon _{33}=\frac{1}{h}w_{|\zeta },
\label{2.7b} \\
E_\alpha &=-\varphi _{,\alpha }, \quad E_3=-\frac{1}{h}\varphi 
_{|\zeta }.
\notag
\end{align}
Here the semicolon preceding Greek indices denotes the co-ordinate expression for the covariant derivatives on the surface, the raising or lowering of indices of surface tensors will be done with the surface metrics $a^{\alpha \beta }$ and $a_{\alpha \beta }$, $b_{\alpha \beta }$ and $c_{\alpha \beta }$ are the second and third fundamental forms of the surface, vertical bar followed by $\zeta $ indicates the partial derivative with respect to $\zeta $ and {\it not} with respect to $x^\zeta $. We denote by $\langle . \rangle $ the integral over $\zeta $ within the limits $[-1/2,1/2]$.

\section{Two-dimensional electro-elastic moduli}

Before applying the variational-asymptotic procedure let us transform the electric enthalpy density to another form more convenient for the asymptotic analysis \citep{Le1986a}. We note that among terms of $W(\varepsilon _{ab},E_a)$ the derivatives $w_{\alpha |\zeta }/h$ and $w_{|\zeta }/h$ in $\varepsilon _{\alpha 3}$ and $\varepsilon _{33}$ as well as $E_3=-\varphi _{|\zeta }/h$ are the main ones in the asymptotic sense. Therefore it is convenient to single out the components $\varepsilon _{\alpha 3}$ and $\varepsilon _{33}$ as well as $E_3$ in the electric enthalpy density. We represent the latter as the sum of two quadratic forms $W_\parallel $ and $W_\perp $ corresponding to longitudinal and transverse electric enthalpy densities, respectively. These are defined by
\begin{align}
W_\parallel &=\min_{\varepsilon _{\alpha 3},
\varepsilon _{33}}\max_{E_3}W, \notag
\\
W_\perp &=W-W_\parallel . \label{2.9}
\end{align}
Let us first find the decomposition \eqref{2.9} in the most general case of anisotropy \citep{Le1986a}. Long, but otherwise simple calculations show that
\begin{align}
W_\parallel &=\frac{1}{2}c^{\alpha \beta \gamma \delta }_N
\varepsilon _{\alpha \beta }\varepsilon _{\gamma \delta }-
e^{\gamma \alpha \beta }_N\varepsilon _{\alpha \beta }E_\gamma 
-\frac{1}{2}\epsilon^{\alpha \beta }_N E_\alpha E_\beta ,
\notag \\
W_\perp &=\frac{1}{2}c^{3333}\gamma ^2+c^{\alpha 333}\gamma
\gamma _\alpha +\frac{1}{2}c^{3\alpha 3\beta }\gamma _\alpha
\gamma _\beta 
\label{2.10} \\
&-e^{333}\gamma F-e^{3\alpha 3}\gamma _\alpha F
-\frac{1}{2}\epsilon^{33} F^2,
\notag
\end{align}
where
\begin{align*}
\gamma &=\varepsilon _{33}+r^{\alpha \beta }\varepsilon _{\alpha \beta 
}-r^\alpha E_\alpha ,
\\
\gamma _\alpha &=2\varepsilon _{\alpha 3}+p_\alpha ^{\mu \nu }
\varepsilon _{\mu \nu }-p_\alpha ^\mu E_\mu ,
\\
F&=E_3+q^{\alpha \beta }\varepsilon _{\alpha \beta }+q^\alpha 
E_\alpha .
\end{align*}
The coefficients $c^{\alpha \beta \gamma \delta }_N$, $e^{\gamma \alpha \beta }_N$, $\epsilon^{\alpha \beta }_N$, $c^{3\alpha 3\beta }$, $c^{\alpha 333}$, $c^{3333}$, $e^{333}$, $e^{3\alpha 3}$, $\epsilon^{33}$, $r^{\alpha \beta }$, $r^\alpha $, $p_\alpha ^{\mu \nu }$, $p_\alpha ^\mu $, $q^{\alpha \beta }$ and $q^\alpha $ can be regarded as components of surface tensors referred to the basis vectors $\mathbf{t}_\alpha $ of the middle surface. We shall call them ``two-dimensional'' electroelastic moduli. They are evaluated in terms of the three-dimensional moduli by means of the formulas
\begin{gather}
c^{\alpha \beta \gamma \delta }_N=c^{\alpha \beta \gamma \delta 
}_P+q^{\alpha \beta }e^{3\gamma \delta }_P, \quad e^{\gamma
\alpha \beta }_N=e^{\gamma \alpha \beta }_P-q^{\alpha \beta }
\epsilon^{\gamma 3}_P,
\notag \\
\epsilon^{\alpha \beta }_N=\epsilon^{\alpha \beta }_P-
q^\alpha \epsilon^{\beta 3}_P, \quad q^{\alpha \beta }=
e^{3\alpha \beta }_P/\epsilon^{33}_P, \quad q^\alpha =
\epsilon^{\alpha 3}_P/\epsilon^{33}_P,
\notag \\
c^{\alpha \beta \gamma \delta }_P=\hat{c}^{\alpha \beta \gamma \delta }
-k^{\alpha \beta }_\nu \hat{c}^{\gamma \delta \nu 3}, \quad
e^{a\alpha \beta }_P=\hat{e}^{a\alpha \beta }-k^{\alpha \beta }_\nu 
\hat{e}^{a\nu 3},
\notag \\
\epsilon^{\alpha b}_P=\hat{\epsilon}^{\alpha b}+k^\alpha _\nu 
\hat{e}^{b\nu 3},\quad \epsilon^{33}_P=\hat{\epsilon}^{33}+k_\nu
\hat{e}^{3\nu 3},
\label{2.11} 
\\
k^{\mu \nu }_\alpha =h_{\alpha \beta }\hat{c}^{\mu \nu \beta 3},\quad
k^\mu _\alpha =h_{\alpha \beta }\hat{e}^{\mu \beta 3},\quad 
k_\alpha =h_{\alpha \beta }\hat{e}^{3\beta 3},\quad 
h_{\alpha \beta }=(\hat{c}^{3\alpha 3\beta })^{-1},
\notag \\
\hat{c}^{a\alpha b\beta }=c^{a\alpha b\beta }-c^{a\alpha 33}
c^{b\beta 33}/c^{3333}, \quad \hat{e}^{ab\beta }=e^{ab\beta }
-c^{b\beta 33}e^{a33}/c^{3333},
\notag \\
\hat{\epsilon}^{ab}=\epsilon^{ab}+e^{a33}e^{b33}/
c^{3333},\quad p^{\mu \nu }_\alpha =k^{\mu \nu }_\alpha +
k_\alpha q^{\mu \nu },
\notag \\
p^\mu _\alpha =k^\mu _\alpha -k_\alpha q^\mu ,\quad 
r^{\alpha \beta } =f^{\alpha \beta }+fq^{\alpha \beta },\quad
r^{\alpha } =f^{\alpha }-fq_{\alpha },
\notag \\
f^{\alpha \beta }=\frac{c^{\alpha \beta 33}-c^{\lambda 333}k^{\alpha
\beta }_\lambda }{c^{3333}},\quad f^\alpha =\frac{e^{\alpha 33}
-c^{\lambda 333}k^\alpha _\lambda }{c^{3333}},\quad 
f =\frac{e^{333}-c^{\lambda 333}k_\lambda }{c^{3333}}.
\notag
\end{gather}

Note that, as these tensors are referred to the basis $\{ \mathbf{t}_\alpha , \mathbf{n}\}$, their components will depend on $\zeta $ through the shifter $\mu ^\beta _\alpha =\delta ^\beta _\alpha -h\zeta b^\beta _\alpha $ even for homogeneous layers. However, since we construct the approximate shell theory admitting the error $h/R$, this weak dependence on $\zeta $ of the 2-D moduli of the homogeneous layers can be neglected. Besides, for our sandwich shell possessing the transversal isotropy all 2-D tensors of odd rank vanish
\begin{displaymath}
e^{\gamma \alpha \beta }_N=c^{\alpha 333}=e^{3\alpha 3}=r^\alpha =p^{\mu \nu }_\alpha=q^{\alpha }=0.
\end{displaymath}
The 2-D tensors of even rank are given by 
\begin{align*}
c^{\alpha \beta \gamma \delta }_N(\zeta )&=\begin{cases}
   c^N_1 a^{\alpha \beta } a^{\gamma \delta }
+c^N_2(a^{\alpha \gamma } a^{\beta \delta }+a^{\alpha \delta } a^{\beta 
\gamma })   & \text{for $|\zeta |>\zeta _0$}, \\
\bar{c}^P_1 a^{\alpha \beta } a^{\gamma \delta }
+\bar{c}^P_2(a^{\alpha \gamma } a^{\beta \delta }+a^{\alpha \delta } a^{\beta 
\gamma })      & \text{for $|\zeta |<\zeta _0$},
\end{cases}
\\
\epsilon^{\alpha \beta }_N(\zeta )&=\epsilon^N(\zeta )a^{\alpha \beta } =\begin{cases}
  \epsilon^Na^{\alpha \beta }   & \text{for $|\zeta |>\zeta _0$}, \\
   \bar{\epsilon}^Na^{\alpha \beta }    & \text{for $|\zeta |<\zeta _0$},
\end{cases}
\end{align*}
\begin{align*}
c^{3333}(\zeta )&=c_{33}(\zeta )=\begin{cases}
  c^E_{33}   & \text{for $|\zeta |>\zeta _0$}, \\
  \bar{c}_{33}    & \text{for $|\zeta |<\zeta _0$},
\end{cases}
\\
c^{3\alpha 3\beta }(\zeta )&=c_{55}(\zeta )a^{\alpha \beta } =\begin{cases}
  c^E_{55} a^{\alpha \beta }   & \text{for $|\zeta |>\zeta _0$}, \\
  \bar{c}_{55} a^{\alpha \beta }   & \text{for $|\zeta |<\zeta _0$},
\end{cases}
\\
e^{333}(\zeta )&=e_{33}(\zeta )=\begin{cases}
  e_{33}   & \text{for $|\zeta |>\zeta _0$}, \\
  0    & \text{for $|\zeta |<\zeta _0$},
\end{cases}
\\
\epsilon^{33}(\zeta )&=\epsilon_{33}(\zeta )=\begin{cases}
  \epsilon^S_{33}   & \text{for $|\zeta |>\zeta _0$}, \\
  \bar{\epsilon}_{33}    & \text{for $|\zeta |<\zeta _0$},
\end{cases}
\\
r^{\alpha \beta }(\zeta )&=r(\zeta )a^{\alpha \beta } =\begin{cases}
  r a^{\alpha \beta }   & \text{for $|\zeta |>\zeta _0$}, \\
  \bar{f} a^{\alpha \beta }  & \text{for $|\zeta |<\zeta _0$},
  \end{cases}
  \\
p^{\alpha \beta }(\zeta )&=p(\zeta )a^{\alpha \beta } =\begin{cases}
  p a^{\alpha \beta }  & \text{for $|\zeta |>\zeta _0$}, \\
  0   & \text{for $|\zeta |<\zeta _0$},
  \end{cases}
  \\
q^{\alpha \beta }(\zeta )&=q(\zeta )a^{\alpha \beta } =\begin{cases}
  q a^{\alpha \beta }  & \text{for $|\zeta |>\zeta _0$}, \\
  0  & \text{for $|\zeta |<\zeta _0$},
\end{cases}
\end{align*}
where $\zeta _0=h_0/2h$ and Voigt's notation is used on the right-hand side of these formulas (see \citep{Le1999}). Taking into account the above properties, we present the longitudinal and transverse electric enthalpy densities in the form
\begin{equation}
\label{2.12}
\begin{split}
W_\parallel =\frac{1}{2}c^N_1(\zeta ) (\varepsilon ^\alpha _\alpha )^2
+c^N_2(\zeta )\varepsilon ^{\alpha \beta }\varepsilon _{\alpha \beta }
-\frac{1}{2}\epsilon^N(\zeta ) E^\alpha E_\alpha ,
\\
W_\perp =\frac{1}{2}c_{33}(\zeta )\gamma ^2+\frac{1}{2}c_{55}(\zeta )a^{\alpha \beta }\gamma _\alpha \gamma _\beta -e_{33}(\zeta )\gamma F-\frac{1}{2}\epsilon_{33}(\zeta ) F^2,
\end{split}
\end{equation}
where
\begin{equation}
\label{2.13}
\begin{split}
\gamma =\varepsilon _{33}+r(\zeta )a^{\alpha \beta }\varepsilon _{\alpha \beta },
\\
\gamma _\alpha =2\varepsilon _{\alpha 3}-p(\zeta )E_\alpha ,
\\
F=E_3+q(\zeta )a^{\alpha \beta }\varepsilon _{\alpha \beta }.
\end{split}
\end{equation}

\section{\bf Asymptotic analysis of the action functional}

We restrict ourselves to the low frequency vibrations of the sandwich shell for which assumption \eqref{2.7a} is valid. Based on this assumption we may neglect the kinetic energy density in the variational-asymptotic procedure.\footnote{For the high-frequency vibrations of elastic and piezoelectric shells and rods where the kinetic energy density should be kept in the variational-asymptotic analysis see \citep{Berdichevsky1980,Berdichevsky1982,Le1985,Le1986b,Le1997,Le1999}.}  
At the first step of the variational-asymptotic procedure we keep only the asymptotically principal terms in the transverse electric enthalpy densities \eqref{2.12} having the formal order $1/h^2$. Since the obtained functional contains only the derivatives with respect to $\zeta $, we drop the integration over $\Omega $ and $t$ and reduce the problem to finding extremal of the functional
\begin{equation}
\label{4.1}
I_0=\int_{-1/2}^{1/2}[\frac{1}{2}c_{33}(\zeta )(w_{|\zeta })^2+\frac{1}{2}c_{55}(\zeta )w^\alpha _{|\zeta }w_{\alpha |\zeta }+e_{33}(\zeta )w_{|\zeta }\varphi _{|\zeta }-\frac{1}{2}\epsilon_{33}(\zeta )(\varphi _{|\zeta })^2]d\zeta 
\end{equation}
among $w_\alpha $, $w$, and $\varphi $ satisfying constraint \eqref{2.1} where the coordinates $x^\alpha $ and time $t$ are regarded as parameters. It is easy to show that the extremal of \eqref{4.1} satisfies the equations
\begin{equation}
\label{4.2}
\begin{split}
(c_{55}(\zeta ) w_{\alpha |\zeta })_{|\zeta }=0,
\\
(c_{33}(\zeta )w_{|\zeta }+e_{33}(\zeta )\varphi _{|\zeta })_{|\zeta }=0,
\\
(e_{33}(\zeta )w_{|\zeta }-\epsilon_{33}(\zeta )\varphi _{|\zeta })_{|\zeta }=0,
\end{split}
\end{equation}
subjected to the boundary conditions
\begin{equation}
\label{4.3}
\begin{split}
c_{55}(\zeta ) w_{\alpha |\zeta }|_{\zeta =\pm 1/2}=0,
\\
(c_{33}(\zeta )w_{|\zeta }+e_{33}(\zeta )\varphi _{|\zeta })|_{\zeta =\pm 1/2}=0,
\\
\varphi (\pm 1/2)=\pm \varphi _0(t)/2,
\end{split}
\end{equation}
the continuity of displacements $w_\alpha$, $w$ and electric potential $\varphi $, as well as the continuity of $c_{55}(\zeta ) w_{\alpha |\zeta }$, $c_{33}(\zeta )w_{|\zeta }+e_{33}(\zeta )\varphi _{|\zeta }$, and $e_{33}(\zeta )w_{|\zeta }-\epsilon_{33}(\zeta )\varphi _{|\zeta }$ when crossing the points $\zeta =-\zeta _0$ and $\zeta =\zeta _0$. Since the functions $c_{55}(\zeta )$, $e_{33}(\zeta )$, and $\epsilon_{33}(\zeta )$ are piecewise constant in the interval $(-1/2,-\zeta _0)$, $(-\zeta _0,\zeta _0)$, and $(\zeta _0,1/2)$, equations \eqref{4.2} admit an exact integration in those intervals. The constants of integrations can be found from the boundary conditions \eqref{4.3} and the continuity conditions. Omitting the calculations, we present the final results 
\begin{align}
w_\alpha (x^\alpha ,\zeta ,t)&=u_\alpha (x^\alpha ,t),\quad w(x^\alpha ,\zeta ,t)=u(x^\alpha ,t)+u^*(\zeta ,t) \notag
\end{align}
\begin{align}
u^*(\zeta ,t)&=\begin{cases}
   0   & \text{for $|\zeta |<\zeta _0$}, \\
   \varphi _0(t)\frac{\epsilon_{33}e_{33}}{c^E_{33}}\frac{\zeta \mp \zeta _0}{\bar{\epsilon}_{33}(2\zeta _0-1)-2\zeta_0\epsilon^S_{33}(1+k_t^2)}  & \text{for $|\zeta |>\zeta _0$},
\end{cases} \label{4.4}
\\
\varphi (x^\alpha ,\zeta ,t)&=\varphi ^*(\zeta ,t)=\begin{cases}
   -\varphi _0(t) \frac{\epsilon^S_{33}(1+k_t^2)}{\bar{\epsilon}_{33}(2\zeta _0-1)-2\zeta _0\epsilon^S_{33}(1+k_t^2)}\zeta  & \text{for $|\zeta |<\zeta _0$}, \\
 \varphi _0(t)\frac{\bar{\epsilon}_{33}\zeta \pm \zeta _0[\bar{\epsilon}_{33}-\epsilon^S_{33}(1+k_t^2)]}{\bar{\epsilon}_{33}(2\zeta _0-1)-2\zeta_0\epsilon^S_{33}(1+k_t^2)}     & \text{for $|\zeta |>\zeta _0$},
\end{cases} \notag
\end{align}
where $u_\alpha $ and $u$ are arbitrary functions of $x^\alpha $ and $t$ and $k_t^2=e^2_{33}/(c^E_{33}\epsilon ^S_{33}$ the thickness coupling factor.

At the second step of the variational-asymptotic procedure we fix $u_\alpha (x^\alpha ,t)$ and $u(x^\alpha ,t)$ and seek the stationary point of the functional \eqref{2.7} in the form
\begin{equation}
\label{4.5}
\begin{split}
w_\alpha (x^\alpha ,\zeta ,t)=u_\alpha (x^\alpha ,t)+hv_\alpha (x^\alpha ,\zeta ,t), 
\\ 
w(x^\alpha ,\zeta ,t)=u(x^\alpha ,t)+u^*(\zeta ,t),
\\
\varphi (x^\alpha ,\zeta ,t)=\varphi ^*(\zeta ,t),
\end{split}
\end{equation}
where $hv_\alpha (x^\alpha ,\zeta ,t)$ is the correction term. By redefining $u_\alpha (x^\alpha ,t)$ if required, we can put the following constraint on the function $v_\alpha (x^\alpha ,\zeta ,t)$:
\begin{displaymath}
\langle v_\alpha (x^\alpha ,\zeta ,t)\rangle =0.
\end{displaymath}
According to the above constraint, $u_\alpha (x^\alpha ,t)$ describes the mean displacements of the shell in the longitudinal directions. Keeping in the electric enthalpy densities \eqref{2.12} only asymptotically principal terms containing $v_\alpha $, we arrive at finding the extremal of the following functional
\begin{displaymath}
I_1=\int_{-1/2}^{1/2}\frac{1}{2}c_{55}(\zeta )a^{\alpha \beta }(v_{\alpha |\zeta }+u_{,\alpha }+b_\alpha ^\lambda u_\lambda )(v_{\alpha |\zeta }+u_{,\alpha }+b_\alpha ^\mu u_\mu )\, d\zeta .
\end{displaymath}
Obviously the extremum is zero and is achieved at $v_{\alpha }=-\varphi _{\alpha }\zeta $, where $\varphi _\alpha =u_{,\alpha }+b_\alpha ^\mu u_\mu $ describe the rotation angles of the middle surface.

At the third step we look for the stationary point of the functional \eqref{2.7} in the form
\begin{equation}
\label{4.6}
\begin{split}
w_\alpha (x^\alpha ,\zeta ,t)=u_\alpha (x^\alpha ,t)-h\zeta \varphi _\alpha +hy_{\alpha }(x^\alpha ,\zeta ,t), 
\\
w(x^\alpha ,\zeta ,t)=u(x^\alpha ,t)+u^*(\zeta ,t)+hy(x^\alpha ,\zeta ,t),
\\
\varphi (x^\alpha ,\zeta ,t)=\varphi ^*(\zeta ,t)+h\chi (x^\alpha ,\zeta ,t),
\end{split}
\end{equation}
where $hy_\alpha $, $hy$, and $h\chi $ are the correction terms. Without restricting generality, we can put the following constraints on the functions $y_\alpha $ and $y$
\begin{equation}\label{4.6a}
\langle y_\alpha (x^\alpha ,\zeta ,t)\rangle =0, \quad \langle y(x^\alpha ,\zeta ,t)\rangle =0.
\end{equation}
With these constraints being fulfilled, we may interpret $u_\alpha $ and $u$ as the mean displacements of the shell. Keeping in the electric enthalpy densities \eqref{2.12} only asymptotically principal terms containing $y_\alpha $, $y$, and $\chi $, we arrive at the problem of finding the extremal of the following functional
\begin{displaymath}
I_2=\int_{-1/2}^{1/2}[\frac{1}{2}c_{33}(\zeta )\gamma ^2+\frac{1}{2}c_{55}(\zeta )\gamma ^\alpha \gamma _\alpha -e_{33}(\zeta )\gamma F-\frac{1}{2}\epsilon_{33}(\zeta )F^2] \, d\zeta ,
\end{displaymath}
where
\begin{equation*}
\begin{split}
\gamma =y_{|\zeta }+r(\zeta )(A^\alpha _\alpha -h\zeta B^\alpha _\alpha ),
\quad
\gamma _\alpha =y_{\alpha |\zeta },
\\
F=-\frac{1}{h}\varphi ^*_{|\zeta }-\chi _{|\zeta }+q(\zeta )(A^\alpha _\alpha -h\zeta B^\alpha _\alpha ),
\end{split}
\end{equation*}
with 
\begin{equation}
\label{4.6b}
\begin{split}
A_{\alpha \beta }=u_{(\alpha ;\beta )}-b_{\alpha \beta }u,
\\
B_{\alpha \beta }=u_{;\alpha \beta }+(u_\lambda b^\lambda _{(\alpha })_{;\beta )}+b^\lambda _{(\alpha }u_{\lambda ;\beta )}-c_{\alpha \beta }u
\end{split}
\end{equation}
describing the measures of extension and bending of the shell middle surface, respectively.

Varying functional $I_2$, we obtain the Euler equations
\begin{equation}
\label{4.7}
\begin{split}
(c_{55}(\zeta ) y_{\alpha |\zeta })_{|\zeta }=\lambda _\alpha ,
\\
(c_{33}(\zeta )\gamma -e_{33}(\zeta )F)_{|\zeta }=\lambda ,
\\
(e_{33}(\zeta )\gamma +\epsilon_{33}(\zeta )F)_{|\zeta }=0,
\end{split}
\end{equation}
(with $\lambda _\alpha $ and $\lambda $ being the Lagrange multipliers that can be found later from the constraints \eqref{4.6a}) subjected to the boundary conditions
\begin{equation}
\label{4.8}
\begin{split}
c_{55}(\zeta ) y_{\alpha |\zeta }|_{\zeta =\pm 1/2}=0,
\\
(c_{33}(\zeta )\gamma -e_{33}(\zeta )F)|_{\zeta =\pm 1/2}=0,
\\
\chi (\pm 1/2)=0,
\end{split}
\end{equation}
the continuity of $y_\alpha $, $y$, and $\chi $, as well as the continuity of $c_{55}(\zeta ) y_{\alpha |\zeta }$, $c_{33}(\zeta )\gamma -e_{33}(\zeta )F$, and $e_{33}(\zeta )\gamma +\epsilon_{33}(\zeta )F$ when crossing the points $\zeta =-\zeta _0$ and $\zeta =\zeta _0$. Equations \eqref{4.7} together with the boundary conditions \eqref{4.8} and the continuity conditions constitute the so-called thickness problem that enables one to find the asymptotic distributions of displacements and electric potential for the piezoelectric sandwich shell. This problem can be solved in the similar manner as that of \eqref{4.2} and \eqref{4.3}. Omitting the long, but otherwise standard integration procedure, we present the final results
\begin{align}
y_\alpha (x_\alpha ,\zeta ,t)&=0,
\\
y(x_\alpha ,\zeta ,t)&=\begin{cases}
   -\bar{f}A^\alpha _\alpha \zeta +\bar{f}hB^\alpha _\alpha \frac{\zeta ^2}{2}+Q_1  & \text{for $|\zeta |<\zeta _0$}, \\
   -rA^\alpha _\alpha \zeta +rhB^\alpha _\alpha \frac{\zeta ^2}{2}+\frac{Q_0k_t^2}{e_{33}(1+k_t^2)} \zeta +Q_3  & \text{for $\zeta >\zeta _0$}, \\
  -rA^\alpha _\alpha \zeta +rhB^\alpha _\alpha \frac{\zeta ^2}{2}+\frac{Q_0k_t^2}{e_{33}(1+k_t^2)} \zeta +Q_4  & \text{for $\zeta <-\zeta _0$}, 
\end{cases} \label{4.9}
\\
\chi (x_\alpha ,\zeta ,t)&=\begin{cases}
 -\frac{\varphi ^*}{h}  -\frac{Q_0}{\bar{\epsilon}_{33}}\zeta +Q_2  & \text{for $|\zeta |<\zeta _0$}, \\
-\frac{\varphi ^*}{h}+ q A^\alpha _\alpha \zeta -qhB^\alpha _\alpha \frac{\zeta ^2}{2}-\frac{Q_0}{\epsilon^P_{33}}\zeta +Q_5     & \text{for $\zeta >\zeta _0$}, \\
-\frac{\varphi ^*}{h}+ q A^\alpha _\alpha \zeta -qhB^\alpha _\alpha \frac{\zeta ^2}{2}-\frac{Q_0}{\epsilon^P_{33}}\zeta +Q_6     & \text{for $\zeta <-\zeta _0$},
\end{cases} \notag
\end{align}
where $\epsilon^P_{33}=\epsilon^S_{33}(1+k_t^2)$. Functions $Q_0$, $\ldots$, $Q_6$ do not depend on $\zeta $, but depend on $x^\alpha $ and $t$ through $A^\alpha _\alpha $, $B^\alpha _\alpha $ and $\varphi _0(t)$. The formulas for them read
\begin{align*}
Q_0&=-\langle \frac{1}{\epsilon^P_{33}(\zeta )} \rangle ^{-1} [\frac{\varphi _0(t)}{h}-qA^\alpha _\alpha (1-2\zeta _0)] ,
\\
Q_1&=-\frac{r}{3}hB^\alpha _\alpha (\frac{1}{8}-\zeta _0^3)+(r-\bar{f})hB^\alpha _\alpha \zeta _0^2(\frac{1}{2}-\zeta _0)-\bar{f}hB^\alpha _\alpha \frac{\zeta _0^3}{3} ,
\\
Q_2&=-qhB^\alpha _\alpha \frac{1}{2}(\zeta _0^2-\frac{1}{4}) ,
\\
Q_3&=(r-\bar{f})A^\alpha _\alpha \zeta _0-\frac{r}{3}hB^\alpha _\alpha (\frac{1}{8}-\zeta _0^3)-(r-\bar{f})hB^\alpha _\alpha \zeta _0^3-\bar{f}hB^\alpha _\alpha \frac{\zeta _0^3}{3}
\\
&+\langle \frac{1}{\epsilon^P_{33}(\zeta )} \rangle ^{-1} [\frac{\varphi _0(t)}{h}-qA^\alpha _\alpha (1-2\zeta _0)]\frac{k_t^2}{e_{33}(1+k_t^2)}\zeta _0 ,
\\
Q_4&=-(r-\bar{f})A^\alpha _\alpha \zeta _0-\frac{r}{3}hB^\alpha _\alpha (\frac{1}{8}-\zeta _0^3)-(r-\bar{f})hB^\alpha _\alpha \zeta _0^3-\bar{f}hB^\alpha _\alpha \frac{\zeta _0^3}{3}
\\
&-\langle \frac{1}{\epsilon^P_{33}(\zeta )} \rangle ^{-1} [\frac{\varphi _0(t)}{h}-qA^\alpha _\alpha (1-2\zeta _0)]\frac{k_t^2}{e_{33}(1+k_t^2)}\zeta _0 ,
\\
Q_5&=\frac{\varphi _0(t)}{2h}-\frac{1}{2}q A^\alpha _\alpha +\frac{1}{8}q hB^\alpha _\alpha +\frac{Q_0}{2\epsilon^P_{33}} ,
\\
Q_6&=-\frac{\varphi _0(t)}{2h}+\frac{1}{2}q A^\alpha _\alpha +\frac{1}{8}q hB^\alpha _\alpha -\frac{Q_0}{2\epsilon^P_{33}} .
\end{align*}
It is interesting to note that both $\gamma $ and $F$ are piecewise constant and are given by
\begin{align}
\label{4.10}
\gamma &=\begin{cases}
  0    & \text{for $|\zeta |<\zeta _0$}, \\
  \frac{Q_0k_t^2}{e_{33}(1+k_t^2)}    & \text{for $|\zeta |>\zeta _0$},
\end{cases}
\\
F&=\begin{cases}
   \frac{Q_0}{\bar{\epsilon}_{33}}   & \text{for $|\zeta |<\zeta _0$}, \\
   \frac{Q_0}{\epsilon^P_{33}}   & \text{for $|\zeta |>\zeta _0$}.
\end{cases}
\end{align}

\section{Two-dimensional theory}

In accordance with the variational-asymptotic method we take now the displacement field and the electric field represented in \eqref{4.6}, where functions $y_\alpha $, $y$, and $\chi $ are given by \eqref{4.9}. We regard $u_\alpha (x^\alpha ,t)$ and $u(x^\alpha ,t)$ as the unknown functions, with $A_{\alpha \beta }$ and $B_{\alpha \beta }$ describing the measures of extension and bending of the shell middle surface, respectively. We substitute this displacement and electric fields into the action functional \eqref{2.7}. Since we construct the approximate theory admitting the error of order $h/R$, $\kappa $ in \eqref{2.7} may be replaced by 1. If we keep only the principal terms containing the unknown functions in the average Lagrangian and integrate over the thickness, then the average kinetic energy density becomes
\begin{equation}
\label{5.1}
\Theta (\dot{u}_\alpha ,\dot{u})=\frac{h}{2}\langle \rho (\zeta )\rangle (a^{\alpha \beta }\dot{w}_\alpha \dot{w}_\beta +\dot{w}^2),
\end{equation}
where
\begin{equation}
\label{5.1a}
\langle \rho (\zeta )\rangle =\rho (1-2\zeta _0)+\bar{\rho } 2\zeta _0.
\end{equation}
To compute the average electric enthalpy density we use the additive decomposition $W=W_\parallel +W_\perp $ that leads to
\begin{equation}
\label{5.2}
\Phi =h\langle W \rangle =h(\langle W_\parallel \rangle + \langle W_\perp \rangle ).
\end{equation}
As $E_\alpha $ is negligibly small on the fields \eqref{4.9}, we may neglect the last term of $W_\parallel $ in \eqref{2.12}$_1$ and approximate the strains $\varepsilon _{\alpha \beta }$ by
\begin{equation}
\label{5.3}
\varepsilon _{\alpha \beta }=w_{(\alpha ;\beta )}-b_{\alpha \beta }w-
h\zeta b^\lambda _{(\alpha }w_{\lambda ;\beta )}+h\zeta 
c_{\alpha \beta }w \approx A_{\alpha \beta }-h\zeta B_{\alpha \beta }.
\end{equation}
Thus, for the average longitudinal electric enthalpy density the integration over the thickness yields
\begin{multline}
\label{5.4}
\langle W_\parallel \rangle =\langle \frac{1}{2}c^N_1(\zeta ) (A^\alpha _\alpha -h\zeta B^\alpha _\alpha )^2
+c^N_2(\zeta )(A^{\alpha \beta }-h\zeta B^{\alpha \beta })(A_{\alpha \beta }-h\zeta B_{\alpha \beta }) \rangle 
\\
=\frac{1}{2}\langle c^N_1(\zeta ) \rangle (A^\alpha _\alpha )^2+\frac{1}{2}h^2\langle \zeta ^2c^N_1(\zeta ) \rangle (B^\alpha _\alpha )^2
+\langle c^N_2(\zeta ) \rangle A^{\alpha \beta }A_{\alpha \beta }+h^2\langle \zeta ^2c^N_2(\zeta ) \rangle B^{\alpha \beta }B_{\alpha \beta },
\end{multline}
where
\begin{equation}
\label{5.4a}
\begin{split}
\langle c^N_1(\zeta ) \rangle =c^N_1(1-2\zeta _0)+\bar{c}_1^P 2\zeta _0, \quad \langle \zeta ^2c^N_1(\zeta ) \rangle =\frac{1}{12}[c^N_1(1-8\zeta _0^3)+\bar{c}_1^P8\zeta _0^3], 
\\
\langle c^N_2(\zeta ) \rangle =c^N_2(1-2\zeta _0)+\bar{c}_2^P 2\zeta _0, \quad \langle \zeta ^2c^N_2(\zeta ) \rangle =\frac{1}{12}[c^N_2(1-8\zeta _0^3)+\bar{c}_2^P8\zeta _0^3].
\end{split}
\end{equation}
Note that the cross terms between $A_{\alpha \beta }$ and $B_{\alpha \beta }$ do not appear in the average electric enthalpy density thank to the symmetric placement of the piezoceramic layers with respect to the middle elastic layer causing the evenness of functions $c^N_1(\zeta )$ and $c^N_2(\zeta )$. If these layers are not placed symmetrically or if only one piezoceramic layer is bonded, the cross terms between $A_{\alpha \beta }$ and $B_{\alpha \beta }$ will certainly appear leading to the cross effects in tension and bending. For the average transverse electric enthalpy density we use the fact that $\gamma _\alpha =0$ while $\gamma $ and $F$ are piecewise constant as described in \eqref{4.10}. Therefore the integral of $W_\perp$ over $\zeta $ can be taken as the sum of three integrals
\begin{equation}
\label{5.5}
\langle W_\perp \rangle =(\frac{1}{2}c^E_{33}\gamma ^2-e_{33}\gamma F-\frac{1}{2}\epsilon^S_{33}F^2)(\int_{-1/2}^{-\zeta _0}d\zeta +\int_{\zeta _0}^{1/2}d\zeta )-\frac{1}{2}\bar{\epsilon}_{33}F^2\int_{-\zeta _0}^{\zeta _0}d\zeta .
\end{equation}
It follows from \eqref{4.10} that, for the piezoceramic layers,
\begin{equation}
\label{5.6}
\frac{1}{2}c^E_{33}\gamma ^2-e_{33}\gamma F-\frac{1}{2}\epsilon^S_{33}F^2=-\frac{Q_0^2}{2\epsilon^P_{33}},
\end{equation}
while for the elastic (dielectric) layer,
\begin{equation}
\label{5.7}
\frac{1}{2}\bar{\epsilon}_{33}F^2=\frac{Q_0^2}{2\bar{\epsilon}_{33}}.
\end{equation}
Thus,
\begin{equation}
\label{5.8}
\langle W_\perp \rangle =-Q_0^2(\frac{\zeta _0}{\bar{\epsilon}_{33}}+\frac{1/2-\zeta _0}{\epsilon^P_{33}})=-\frac{1}{2}\langle \frac{1}{\epsilon^P_{33}(\zeta )} \rangle ^{-1} [\frac{\varphi _0(t)}{h}-qA^\alpha _\alpha (1-2\zeta _0)]^2,
\end{equation}
because for the dielectric material $\bar{\epsilon}_{33}=\bar{\epsilon}^P_{33}$. Combining the average longitudinal and transverse electric enthalpy densities together, we obtain 
\begin{multline}
\label{5.9}
\Phi (A_{\alpha \beta },B_{\alpha \beta })=\frac{h}{2}\langle c^N_1(\zeta ) \rangle (A^\alpha _\alpha )^2+\frac{1}{2}h^3\langle \zeta ^2c^N_1(\zeta ) \rangle (B^\alpha _\alpha )^2
+h\langle c^N_2(\zeta ) \rangle A^{\alpha \beta }A_{\alpha \beta }
\\
+h^3\langle \zeta ^2c^N_2(\zeta ) \rangle B^{\alpha \beta }B_{\alpha \beta }-\frac{h}{2}\langle \frac{1}{\epsilon^P_{33}(\zeta )} \rangle ^{-1} [\frac{\varphi _0(t)}{h}-qA^\alpha _\alpha (1-2\zeta _0)]^2.
\end{multline}

We formulate now the variational principle for the smart sandwich shell: the average displacement field $\mathbf{u}(x^\alpha ,t)$ of the sandwich shell changes in space and time in such a way that the 2-D average action functional
\begin{equation}
\label{5.10}
J[\mathbf{u}(x^\alpha ,t)]=\int_{t_0}^{t_1}\int_{\Omega }[\Theta (\dot{\mathbf{u}})-\Phi (A_{\alpha \beta },B_{\alpha \beta })]\, da \, dt
\end{equation} 
becomes stationary among all continuously differentiable functions $\mathbf{u}(x^\alpha ,t)$  satisfying the initial and end conditions
\begin{displaymath}
\mathbf{u}(x^\alpha ,t_0)=\mathbf{u}_0(x^\alpha ), \quad \mathbf{u}(x^\alpha ,t_1)=\mathbf{u}_1(x^\alpha ).
\end{displaymath}
The standard calculus of variation shows that the stationarity condition $\delta J=0$ implies the following two-dimensional equations
\begin{equation}
\langle \rho \rangle \ddot{u}^\alpha =T^{\alpha \beta }_{;
\beta }+b^\alpha _\lambda M^{\lambda \beta }_{;\beta } 
 ,
\label{5.11}
\end{equation}
and
\begin{equation}
\langle \rho \rangle \ddot{u}=T^{\alpha \beta }b_{\alpha \beta }
-M^{\alpha \beta }_{;\alpha \beta },
\label{5.12}
\end{equation}
subjected to the free-edge boundary conditions
\begin{gather}
T^{\alpha \beta }\nu _{\beta }+b^\alpha _\gamma 
M^{\gamma \beta } \nu _\beta =0, \notag \\ 
M^{\alpha \beta }_{;\alpha }\nu _\beta +\frac{\partial }{\partial s}(M^{\alpha \beta }
\tau _\alpha \nu _\beta )=0,
\label{5.13} 
\\
M^{\alpha \beta }\nu _\alpha \nu _\beta =0,
\notag
\end{gather}
where $T^{\alpha \beta }=N^{\alpha \beta }+b^\alpha _\lambda M^{\lambda \beta }$ and $\nu _\alpha $ denotes the components of the surface vector normal to the curve $\partial \Omega$. For the clamped or simply supported edge, natural boundary conditions must be replaced by the corresponding kinematical boundary conditions. The equations of motion \eqref{5.11} and \eqref{5.12} must be complemented by the constitutive equations
\begin{align}
N^{\alpha \beta }&= \frac{\partial \Phi }{\partial A_{\alpha \beta }}=h\langle c^N_1(\zeta )\rangle  
A^\lambda _\lambda a^{\alpha \beta }+2h\langle c^N_2(\zeta )\rangle  
A^{\alpha \beta }
\notag \\
&+\langle \frac{1}{\epsilon^P_{33}(\zeta )} \rangle ^{-1}[\varphi _0(t)-qA^\lambda _\lambda (1-2\zeta _0)]a^{\alpha \beta } \label{5.14}
\\
M^{\alpha \beta }&= \frac{\partial \Phi }{\partial B_{\alpha \beta }}=h^3\langle \zeta ^2c^N_1(\zeta )\rangle  
B^\lambda _\lambda a^{\alpha \beta }+2h^3\langle \zeta ^2c^N_2(\zeta )\rangle  
B^{\alpha \beta }.
\end{align}
Note that, if the elastic layer is not fully covered by the piezoelectric patches, the average mass density and stiffnesses of the 2-D theory will suffer jump across the boundary between the covered and uncovered area. In this case the jump conditions along this boundary can also be obtained (see Section 7). It is readily seen that the constructed two-dimensional theory of sandwich shells reduces to the classical theories of elastic shells (with $\zeta _0=1/2$) and piezoelectric shells (with $\zeta _0=0$), respectively (cf. \citep{Le1999}).

To complete the 2-D theory of piezoelectric sandwich shells we should also indicate the method of restoring the 3-D electroelastic state by means of the 2-D one. To do this, the strain tensor $\boldsymbol{\varepsilon}$ and the electric field $\mathbf{E}$ should be found from \eqref{2.7b}. It  can be shown by the asymptotic analysis that the following formulas
\begin{equation}
\begin{split}
\varepsilon _{\alpha \beta }=A_{\alpha \beta }- h\zeta B_{
\alpha \beta },\quad 2\varepsilon _{\alpha 3}=
y_{\alpha |\zeta },\quad \varepsilon _{33}=y_{|\zeta },
\\
E_\alpha =0 ,\quad E_3=-\frac{1}{h}\varphi ^*_{|\zeta }-\chi _{|\zeta }
\end{split}
\label{5.15}
\end{equation}
hold true within the first-order approximation. Using \eqref{4.6}, we find that
\begin{align}
\varepsilon _{\alpha \beta }&=A_{\alpha \beta }- h\zeta B_{
\alpha \beta },
\quad
2\varepsilon _{\alpha 3}= 0, \notag
\\
\varepsilon _{33}&=\begin{cases}
   -\bar{f}A^\alpha _\alpha +\bar{f}hB^\alpha _\alpha \zeta  & \text{for $|\zeta |<\zeta _0$}, \\
   -rA^\alpha _\alpha +rhB^\alpha _\alpha \zeta +\frac{Q_0k_t^2}{e_{33}(1+k_t^2)} & \text{for $|\zeta |>\zeta _0$}, 
\end{cases}, \label{5.15a}
\\
E_\alpha &=0 ,\quad E_3=\begin{cases}
 \frac{Q_0}{\bar{\epsilon}_{33}} & \text{for $|\zeta |<\zeta _0$}, \\
 -q A^\alpha _\alpha +qhB^\alpha _\alpha \zeta +\frac{Q_0}{\epsilon^P_{33}}   & \text{for $|\zeta |>\zeta _0$}
 \end{cases}. \notag
\end{align}
Note that, in contrast to the homogeneous piezoelectric shells, the strain $\varepsilon _{33}$ and the electric field $E_3$ in the smart sandwich shells are discontinuous through the thickness. The stress tensor $\boldsymbol{\sigma }$ and the electric induction $\mathbf{D}$ are then determined by the 3-D constitutive equations. While doing so, it is convenient to use the decomposition \eqref{2.9} for the electric enthalpy density. Within the first-order approximation we find \begin{gather}
\sigma^{\alpha \beta }=\frac{1}{h}N^{\alpha \beta }-\frac{12}{h^2}M^{\alpha \beta }\zeta ,
\quad
\sigma^{\alpha 3}=0,\quad \sigma^{33}=0,
\label{5.16} 
\\
D^\alpha =0 ,\quad
D^3=Q_0=-\langle \frac{1}{\epsilon^P_{33}(\zeta )} \rangle ^{-1} [\frac{\varphi _0(t)}{h}-qA^\alpha _\alpha (1-2\zeta _0)].
\notag
\end{gather}
Note that $\boldsymbol{\sigma }$ and $\mathbf{D}$ are continuous through the thickness. Again, these formulae are accurate up to terms of the orders $h/R$ and $h/l$ of smallness.

\section{Error estimation of the constructed 2-D theory}
In this Section we shall prove an identity that generalizes Prager-Synge's identity found in \citep{Prager1947} to the statics of inhomogeneous piezoelectric bodies. Based on this identity an error estimate of the smart sandwich shell theory constructed in the previous Section for the special case of statics will be established. 

We consider an inhomogeneous piezoelectric body occupying the three-dimensional domain $\mathcal{V}$ in its undeformed state that stays in equilibrium under a fixed voltage. Concerning the boundary conditions for the mechanical quantities we assume that the boundary $\partial \mathcal{V}$ is decomposed into two subboundaries $\partial _k$ and $\partial_s$. On the part $\partial_k$ the displacements vanish (clamped boundary)
\begin{equation}
\mathbf{w}=0 \quad \text{on $\partial_k$}.
\label{6.a}
\end{equation}
On the remaining part $\partial_s$ the traction-free boundary condition is assumed
\begin{equation}
\boldsymbol{\sigma }\cdot \mathbf{n} = \mathbf{0} \quad \text{on
$\partial_s$}.
\label{6.b}
\end{equation}
Concerning the boundary conditions for the electric potential we assume that the boundary $\partial \mathcal{V}$ consists of $n+1$ subboundaries $\partial_e^{(1)}, \ldots , \partial_e^{(n)}$ and $ \partial_d$. The subboundaries $\partial_e^{(1)},\ldots , \partial_e^{(n)}$ are covered by electrodes with negligible thickness. On these electrodes the electric potential is prescribed
\begin{equation}
\varphi =\varphi _{(i)} \quad \text{on
$\partial_e^{(i)},i=1,\ldots ,n$}.
\label{6.c}
\end{equation}
On the uncoated portion $\partial_d$ of the boundary we require that the surface charge vanishes
\begin{equation}
\mathbf{D}\cdot \mathbf{n} = 0 \quad \text{on $\partial_d$}.
\label{6.d}
\end{equation}

We introduce the linear vector space of electroelastic states that consists of elements of the 
form $\boldsymbol{\Xi} =(\boldsymbol{\sigma },\mathbf{E})$, where $\boldsymbol{\sigma }$ is the stress field and $\mathbf{E}$ is the electric field; both fields are defined in the three-dimensional domain $\mathcal{V}$ occupied by the piezoelectric body. In this space we introduce the following energetic norm
\begin{equation}
\parallel \boldsymbol{\Xi} \parallel ^2_{L_2}=C_2[\boldsymbol{\Xi} ]=\int_{\mathcal{V}}
G(\boldsymbol{\sigma },\mathbf{E})\, dv,
\label{6.1}
\end{equation}
where function $G(\boldsymbol{\sigma },\mathbf{E})$ is the density of the complementary energy (or Gibbs function) \citep{Le1999}. In component form $G(\boldsymbol{\sigma },\mathbf{E})$ reads
\begin{eqnarray*}
G(\boldsymbol{\sigma },\mathbf{E})=\frac{1}{2}s^E_{abcd}\sigma^{ab}\sigma^{cd}+
d_{cab}\sigma^{ab}E^c+\frac{1}{2}\epsilon ^T_{ab}E^aE^b,
\end{eqnarray*}
where the precise dependence of the electroelastic moduli on $\mathbf{x}$ is suppressed for short. Since the complementary energy density $G(\boldsymbol{\sigma },\mathbf{E})$ is positive definite, the definition \eqref{6.1} is meaningful. 

We call ``kinematically admissible'' those electroelastic states $\check{\boldsymbol{\Xi} }$ for which the compatible strain field $\check{\boldsymbol{\varepsilon }}$ and the electric induction field $\check{\mathbf{D}}$ exist such that
\begin{gather*}
\check{\boldsymbol{\varepsilon }}=\frac{1}{2}(\nabla \check{\mathbf{w}}+(\nabla 
\check{\mathbf{w}})^T), \quad \check{\mathbf{w}}=0 \quad 
\text{on $\partial_k$},
\\
\text{div}\check{\mathbf{D}}=0,\quad
\check{\mathbf{D}}\cdot \mathbf{n}=0 \quad \text{on $\partial_d$},
\end{gather*}
while $\check{\boldsymbol{\sigma }}$ and $\check{\mathbf{E}}$ are expressed in terms of $\check{\boldsymbol{\varepsilon }}$ and $\check{\mathbf{D}}$ by the constitutive equations equivalent to \eqref{2.6}. We call those electroelastic states $\hat{\boldsymbol{\Xi} }$ ``statically admissible'', when
\begin{gather*}
\text{div}\hat{\boldsymbol{\sigma }}=0,\quad \hat{\boldsymbol{\sigma }}\cdot \mathbf{n}=0 \quad
\text{on $\partial_s$},
\\
\hat{\mathbf{E}}=-\nabla \hat{\varphi }, \quad \hat{\varphi 
}=\varphi _{(i)} \quad \text{on $\partial _e^{(i)},i=1,\ldots 
,n$}.
\end{gather*}

Let $\tilde{\boldsymbol{\Xi} }=(\tilde{\boldsymbol{\sigma }},\tilde{\mathbf{E}})$ be the true 
electroelastic state that is realized in an inhomogeneous piezoelectric body staying in equilibrium under the given values of the electric potential $\varphi _{(i)}$ on the electrodes $\partial _e^{(i)},i=1,\ldots ,n$. Then the following identity
\begin{equation}
C_2[\tilde{\boldsymbol{\Xi}}-\frac{1}{2}(\check{\boldsymbol{\Xi} }+\hat{\boldsymbol{\Xi}})]=
C_2[\frac{1}{2}(\check{\boldsymbol{\Xi}}-\hat{\boldsymbol{\Xi}})]
\label{6.2}
\end{equation}
turns out to be valid for arbitrary kinematically and statically admissible fields $\check{\boldsymbol{\Xi} }$ and $\hat{\boldsymbol{\Xi} }$. This identity generalizes the well-known Prager-Synge identity \citep{Prager1947} to the statics of inhomogeneous piezoelectric bodies. It implies that $\frac{1}{2}(\check{\boldsymbol{\Xi} }+\hat{\boldsymbol{\Xi} })$ may be regarded as an ``approximation'' to the true solution in the energetic norm, if the complementary energy associated with the difference $\frac{1}{2}(\check{\boldsymbol{\Xi} }-\hat{\boldsymbol{\Xi} })$ is ``small''. In this case we may also consider each of the fields $\check{\boldsymbol{\Xi} }$ or $\hat{\boldsymbol{\Xi} }$ as an ``approximation'', in view of the inequalities
\begin{gather*}
C_2[\tilde{\boldsymbol{\Xi} }-\check{\boldsymbol{\Xi} })]\le C_2[\check{\boldsymbol{\Xi} }-\hat{\boldsymbol{\Xi} }],
\\
C_2[\tilde{\boldsymbol{\Xi} }-\hat{\boldsymbol{\Xi} })]\le C_2[\check{\boldsymbol{\Xi} }-\hat{\boldsymbol{\Xi} }],
\end{gather*}
which follow easily from \eqref{6.2}.

To prove the identity \eqref{6.2} we first rewrite its left-hand side as follows
\begin{align}
C_2[\tilde{\boldsymbol{\Xi} }-\frac{1}{2}(\check{\boldsymbol{\Xi} }+\hat{\boldsymbol{\Xi} })]&= 
C_2[\tilde{\boldsymbol{\Xi} }-\hat{\boldsymbol{\Xi} }-\frac{1}{2}(\check{\boldsymbol{\Xi} }-\hat{\boldsymbol{\Xi} })]
\notag \\
&=C_2[\tilde{\boldsymbol{\Xi} }-\hat{\boldsymbol{\Xi} }]+C_2[\frac{1}{2}(\check{\boldsymbol{\Xi} }
-\hat{\boldsymbol{\Xi} })]-[\tilde{\boldsymbol{\Xi} }-\hat{\boldsymbol{\Xi} },\check{\boldsymbol{\Xi} }-\hat{\boldsymbol{\Xi} }]
\notag \\
&=C_2[\frac{1}{2}(\check{\boldsymbol{\Xi} }
-\hat{\boldsymbol{\Xi} })]+[\tilde{\boldsymbol{\Xi} }-\hat{\boldsymbol{\Xi} },\tilde{\boldsymbol{\Xi} }-\check{\boldsymbol{\Xi} 
}],
\label{6.3}
\end{align}
where $[\boldsymbol{\Xi} ,\boldsymbol{\Xi} ^\prime ]$ denotes the scalar product of two
elements
\begin{equation}
[\boldsymbol{\Xi} ,\boldsymbol{\Xi} ^\prime ]=\int_{\mathcal{B}}G(\boldsymbol{\Xi} ,\boldsymbol{\Xi} ^\prime )\, dv
=\int_{\mathcal{B}}s_{AB}\boldsymbol{\Xi} ^A\boldsymbol{\Xi} ^{\prime B}\, dv.
\label{6.4}
\end{equation}
In \eqref{6.4} $\boldsymbol{\Xi} ^A=(\boldsymbol{\sigma }^{\mathfrak{n}},E^b)$ and $s_{AB}$ is a symmetric matrix, whose elements are themselves matrices
\begin{equation}
s_{AB}=\begin{pmatrix}s^E_{\mathfrak{m} \mathfrak{n}} & 
(d)^T_{\mathfrak{m} b}\\
d_{a\mathfrak{n}} & \epsilon ^T_{ab}\end{pmatrix}=\begin{pmatrix}
c_D^{\mathfrak{m} \mathfrak{n}} & -(h)^{T\mathfrak{m}b}\\
-h^{a\mathfrak{n}} & \beta _S^{ab}\end{pmatrix}^{-1},
\label{6.5}
\end{equation}
where $(d)^T_{\mathfrak{m} b}=d_{b\mathfrak{m}}$ are elements of the transpose matrix and Voigt's abbreviated index notation is used. According to \eqref{6.3}, the identity \eqref{6.2} holds true, when
\begin{eqnarray*}
[\tilde{\boldsymbol{\Xi} }-\hat{\boldsymbol{\Xi} },\tilde{\boldsymbol{\Xi} }-\check{\boldsymbol{\Xi} }]=0.
\end{eqnarray*}
This identity follows from the definitions of $\tilde{\boldsymbol{\Xi} }$, $\check{\boldsymbol{\Xi} }$, $\hat{\boldsymbol{\Xi} }$ and the formulae \eqref{6.4} and \eqref{6.5}. Indeed
\begin{align*}
[\tilde{\boldsymbol{\Xi} }-\hat{\boldsymbol{\Xi} },\tilde{\boldsymbol{\Xi} }-\check{\boldsymbol{\Xi} }]&=
\int_{\mathcal{V}}[(\tilde{\boldsymbol{\sigma }}-\hat{\boldsymbol{\sigma }})\mathbf{:}
(\tilde{\boldsymbol{\varepsilon }}-\check{\boldsymbol{\varepsilon }})+(\tilde{\mathbf{E}}-\hat{\mathbf{E}})
\cdot (\tilde{\mathbf{D}}-\check{\mathbf{D}})]\, dv
\\
&=\int_{\mathcal{V}}[(\tilde{\boldsymbol{\sigma }}-\hat{\boldsymbol{\sigma }})\mathbf{:}
\nabla (\tilde{\mathbf{w}}-\check{\mathbf{w}})+\nabla (\tilde{\varphi }
-\hat{\varphi })
\cdot (\tilde{\mathbf{D}}-\check{\mathbf{D}})]\, dv ,
\end{align*}
which is the consequence of the definitions of $\tilde{\boldsymbol{\Xi} }$, $\check{\boldsymbol{\Xi} }$ and \eqref{6.5}. Integrating this identity by parts and taking the definition of $\hat{\boldsymbol{\Xi} }$ as well as the boundary conditions into account, we see that the right-hand side vanishes. Thus, the identity \eqref{6.2} is proved.

Based on \eqref{6.2} the following error estimate can be established.\footnote{This error estimation generalizes the results obtained first by \citet{Koiter1970} for the elastic shells and by \citet{Le1986a} for the homogeneous piezoelectric shells.}

{\it Theorem.} The electroelastic state determined by the 2-D static theory of piezoelectric sandwich shells differs in the norm $L_2$ from the exact electroelastic state determined by the 3-D theory of piezoelectricity by a quantity of the order $h/R+h/l$ as compared with unity.

To prove this theorem it is enough to find out the kinematically and statically admissible 3-D fields of electroelastic states that differ from the electroelastic state determined by the 2-D theory by a quantity of the order $h/R+h/l$ as compared with unity. Below we shall construct these 
fields.

{\it Construction of kinematically admissible field.} We specify the kinematically admissible 
displacement field in the form
\begin{align*}
\check{w}_\alpha &=u_\alpha (x^\alpha )-x \varphi _{\alpha } 
(x^\alpha ) +hy_\alpha (x^\alpha ,x),
\\
\check{w}&=u(x^\alpha )+u^*(x)+hy (x^\alpha ,x),
\end{align*}
where $\varphi _\alpha =u_{,\alpha }+b_\alpha ^\mu u_\mu $, while $u^*(x)$, $y_\alpha (x^\alpha ,x)$, $y(x^\alpha ,x)$ are given by \eqref{4.4} and \eqref{4.9}, respectively. Here and below, all quantities without the superscripts $\hat{\null }$ and $\check{\null }$ refer to the solution of the equilibrium equations of piezoelectric sandwich shells obtained by the constructed two-dimensional theory. The components of the strain tensor are calculated according to \eqref{2.7b}. Assume that the 2-D electroelastic state is characterized by the strain amplitude $\varepsilon 
=\varepsilon _A+\varepsilon _B$. The asymptotic analysis similar to that given in Section 4 shows that
\begin{align*}
\check{\varepsilon }_{\alpha \beta }&=A_{\alpha \beta }-xB_{\alpha 
\beta }+O(h/R,h/l)\varepsilon =\varepsilon _{\alpha \beta }+
O(h/R,h/l)\varepsilon ,
\notag \\
2\check{\varepsilon }_{\alpha 3}&= O(h/R,h/l)\varepsilon ,
\quad
\check{\varepsilon }_{33}=\varepsilon _{33},
\notag
\end{align*}
with $\varepsilon _{33}$ from \eqref{5.15a}. We choose the components $\check{D}^\alpha $ of the electric induction to be zero, while
\begin{equation}
\check{D}^3=Q_0/\kappa (x).
\end{equation}
It is easy to see that $\check{\mathbf{D}}$ satisfies the exact 3-D equation of electrostatics
\begin{equation}
(\check{D}^\alpha \kappa )_{;\alpha }+(\check{D}^3\kappa 
)_{,x}=0.
\label{6.6}
\end{equation}
and, due to the property of $\kappa $,
\begin{eqnarray*}
\check{D}^3=Q_0+O(h/R,h/l)\epsilon .
\end{eqnarray*}
Note that the constructed field $\check{\mathbf{D}}$ does not satisfy the exact boundary condition $\check{D}^\alpha \kappa \nu _\alpha =0$, posed at the portion $\partial_d\times [-h/2,h/2]$ of the edge. For simplicity of the proof we further assume that the 3-D boundary conditions at the edge of the shell agree with the inner expansion of the electro\-elastic state (the so-called regular boundary conditions). Then the electric induction field $\check{\bf D}$ constructed above is kinematically admissible. For irregular boundary conditions we have to take into account an additional electric induction field that differs substantially from zero only in a thin boundary layer at the shell edge. Since the energy of this boundary layer is of the order $h/l$ compared with that of the inner domain, one can easily generalize the proof of the theorem to this case.

Knowing $(\check{\boldsymbol{\varepsilon }},\check{\mathbf{D}})$, we find $\check{\boldsymbol{\Xi} }=(\check{\boldsymbol{\sigma }},\check{\mathbf{E}})$ from the constitutive equations equivalent to \eqref{2.6}. Because $(\check{\boldsymbol{\varepsilon }},\check{\mathbf{D}})=(\boldsymbol{\varepsilon },\mathbf{D})+O(h/R,h/l)\varepsilon $, it is easily seen that $(\check{\boldsymbol{\sigma }},\check{\mathbf{E}})=(\boldsymbol{\sigma },\mathbf{E})+O(h/R,h/l)\varepsilon$.

{\it Construction of statically admissible field.} We write down the exact 3-D equilibrium equations for a shell in the form (cf. \citep{Le1999})
\begin{equation}
\begin{split}
\hat{\tau }^{\alpha \beta }_{;\beta }
+(\mu ^\alpha _\beta \hat{\tau }^\beta )_{,x}-\hat{\tau }^\beta b^\alpha 
_\beta =0,
\\
\hat{\tau }^\beta _{;\beta }+\hat{\tau }^{\alpha 
\beta }b_{\alpha \beta }+\hat{\tau }_{,x}=0,
\end{split}
\label{6.7}
\end{equation}
where
$$\hat{\tau }^{\alpha \beta }=\mu ^\alpha _\lambda \hat{\sigma }
^{\lambda \beta }\kappa ,\quad \hat{\tau }^\alpha =\hat{\sigma }^{\alpha 3}
\kappa ,\quad \hat{\tau }=\hat{\sigma }^{33}\kappa .$$
Note that $\hat{\tau }^{\alpha \beta }$ is unsymmetric. To find the statically admissible stress field $\hat{\boldsymbol{\sigma }}$ satisfying \eqref{6.7} and the traction-free boundary conditions
\begin{equation}
\mu ^\alpha _\beta \hat{\sigma }^{\beta 3}\kappa =0,\quad
\hat{\sigma }^{33}\kappa =0 \quad \text{at $x=\pm h/2$,}
\label{6.8}
\end{equation}
we proceed as follows. We specify $\hat{\sigma}^{\alpha \beta }$ in 
the form
\begin{eqnarray*}
\hat{\sigma}^{\alpha \beta }=s^{\alpha \beta }_0(x^\alpha )-
xs^{\alpha \beta }_1(x^\alpha ),
\end{eqnarray*}
where $s^{\alpha \beta }_0$ and $s^{\alpha \beta }_1$ are symmetric and independent of $x$. These are chosen from the conditions
\begin{equation}
\langle \hat{\tau }^{\alpha \beta }\rangle _x=T^{\alpha \beta },
\quad \langle \hat{\tau }^{\alpha \beta }x\rangle _x=-M^{\alpha 
\beta },
\label{6.9}
\end{equation}
where $\langle . \rangle _x=\int . dx/h$. The conditions \eqref{6.9} enable one to determine $s^{\alpha \beta }_0$ and $s^{\alpha \beta }_1$ through $T^{\alpha \beta }$ and $M^{\alpha \beta }$ uniquely. Moreover, one can check that
\begin{align*}
s^{\alpha \beta }_0&=\frac{1}{h}N^{\alpha \beta }+O(h/R,h/l)\varepsilon,
\\
s^{\alpha \beta }_1&=\frac{12}{h^3}M^{\alpha \beta }+O(h/R,h/l)\varepsilon .
\end{align*}
Solving \eqref{6.7},\eqref{6.8} with the given $\hat{\tau }^{\alpha \beta }$, we can find $\hat{\tau }^\alpha $ and $\hat{\tau }$ and then $\hat{\sigma}^{\alpha 3}$ and $\hat{\sigma }^{33}$. It turns out that \eqref{6.9} are the sufficient conditions for the existence of $\hat{\tau }^\alpha $ and $\hat{\tau }$. Indeed, integrating \eqref{6.7} and \eqref{6.7}$_1$ multiplied by $x$ over
$x\in [-h/2,h/2]$, we obtain
\begin{equation}
\begin{split}
T^{\alpha \beta }_{;\beta }-b^\alpha _\beta N^\beta +(\mu ^\alpha _\beta 
\hat{\tau }^\beta )|_{-h/2}^{h/2}=0,
\\
N^\alpha _{;\alpha }+b_{\alpha \beta }T^{\alpha \beta }+\hat{\tau }|_
{-h/2}^{h/2}=0,
\\
-M^{\alpha \beta }_{;\beta }-N^\alpha +(x\mu ^\alpha _\beta 
\hat{\tau }^\beta )|_{-h/2}^{h/2}=0,
\end{split}
\label{6.10}
\end{equation}
where $N^\alpha =\langle \hat{\tau }^\alpha \rangle _x$. From the first and the last equations of \eqref{6.10} it follows that $(\mu ^\alpha _\beta \hat{\tau }^\beta )|^{h/2}_{-h/2}=0$, since
$T^{\alpha \beta }_{;\beta }+b^\alpha _\lambda M^{\lambda \beta }_{;\beta }=0$ according to the 2-D equations of equilibrium. From the second equation of \eqref{6.10} we also obtain 
$\hat{\tau }|^{h/2}_{-h/2}=0$. Thus, if the boundary conditions \eqref{6.8} are satisfied at $x=-h/2$, then after the integration they will also be satisfied at $x=h/2$. Not showing the cumbersome solution of \eqref{6.7}, we note only that $\hat{\sigma}^{\alpha 3}, \hat{\sigma}^{33}\sim O(h/R,h/l)\varepsilon$. Thus, $\hat{\boldsymbol{\sigma }}=\boldsymbol{\sigma }
+O(h/R,h/l)\varepsilon $.

Concerning the statically admissible electric field $\hat{\mathbf{E}}$ we specify its potential by
\begin{eqnarray*}
\hat{\varphi }(x^\alpha ,x)=\varphi ^*(x)+h\chi (x^\alpha ,x),
\end{eqnarray*}
with $\varphi ^*(x)$ and $\chi (x^\alpha ,x)$ from \eqref{4.4} and \eqref{4.9}, respectively. Then
\begin{align*}
\hat{E}_3&= \begin{cases}
 \frac{Q_0}{\bar{\epsilon}_{33}} & \text{for $|\zeta |<\zeta _0$}, \\
 -q A^\alpha _\alpha +qhB^\alpha _\alpha \zeta +\frac{Q_0}{\epsilon^P_{33}}   & \text{for $|\zeta |>\zeta _0$}
 \end{cases},
\\
\hat{E}_\alpha &=O(h/l)\varepsilon .
\end{align*}
Note that the statically admissible field $(\hat{\boldsymbol{\sigma }},\hat{\mathbf{E}})$ constructed above satisfies only the regular boundary conditions at the shell edge, exactly as in the previous case.

From this construction we see that $\check{\boldsymbol{\Xi} }$ and $\hat{\boldsymbol{\Xi} }$ differ from those found by the 2-D theory by a quantity of the order $h/R$ and $h/l$ as compared with unity. We have thus established the asymptotic accuracy of the 2-D theory in the energetic norm \eqref{6.1}.

\section{Frequency spectra of circular smart sandwich plates}

In this Section we illustrate the application of the theory to the problem of axisymmetric longitudinal vibration of an elastic circular plate of radius $r$ partially covered by the piezoceramic patches with thickness polarization. Suppose that the piezoceramic patches cover only the rings $\Omega _1$ (defined by $r_0\le \varrho \le r$) of the face surfaces, where $\varrho ,\phi $ are the polar co-ordinates. When these piezoceramic patches are subjected to an oscillating voltage, the electric field occurs leading to the extension or contraction of the patches and forcing the plate to vibrate in the radial direction. For the axisymmetric longitudinal vibration of the plate the bending measures vanish, while for the measures of extension we have
\begin{equation*}
A_{\varrho \varrho }=u_{\varrho ,\varrho },\quad A_{\varrho \phi }=0,
\quad A_{\phi \phi }=\frac{u_\varrho }{\varrho },
\end{equation*}
where $u_\varrho (\varrho ,t)$ is the radial displacement. Thus, the action functional, up to an unimportant factor $\pi $, becomes the sum of the following integrals
\begin{gather}
\int_{t_0}^{t_1}\int_0^{r_0} \bar{\rho } h_0\dot{u}_\varrho ^2 \varrho d\varrho \, dt 
-\int_{t_0}^{t_1}\int_0^{r_0}h_0[\bar{c}^P_1(u_{\varrho 
,\varrho }+\frac{u_\varrho }{\varrho })^2+2\bar{c}^P_2u_{\varrho ,\varrho }^2
+2\bar{c}^P_2\frac{u_\varrho ^2}{\varrho ^2}]\varrho d\varrho \, dt \notag
\\
+\int_{t_0}^{t_1}\int_{r_0}^r \rho ^L h\dot{u}_\varrho ^2 \varrho d\varrho \, dt-\int_{t_0}^{t_1}\int_{r_0}^r h[c_1^L(u_{\varrho 
,\varrho }+\frac{u_\varrho }{\varrho })^2+2c_2^Lu_{\varrho ,\varrho }^2
+2c_2^P\frac{u_\varrho ^2}{\varrho ^2} \notag
\\
+2e^L\frac{\varphi _0(t)}{h}(u_{\varrho ,\varrho }+\frac{u_\varrho }{\varrho }) ]\varrho d\varrho \, 
dt. \label{7.0}
\end{gather}
We use label $L$ to indicate the coefficients in this functional for the sandwich plate with three layers that are functions of $\zeta _0$
\begin{align*}
\rho ^L(\zeta _0)&=\langle \rho (\zeta )\rangle =\rho (1-2\zeta _0)+\bar{\rho } 2\zeta _0,
\\
c_1^L(\zeta _0)&=\langle c^N_1(\zeta ) \rangle -\langle \frac{1}{\epsilon^P_{33}(\zeta )} \rangle ^{-1} q^2(1-2\zeta _0)^2
\\
&=c^N_1(1-2\zeta _0)+\bar{c}_1^P 2\zeta _0-\langle \frac{1}{\epsilon^P_{33}(\zeta )} \rangle ^{-1} q^2(1-2\zeta _0)^2,
\end{align*}
\begin{align*}
c_2^L(\zeta _0)&=\langle c^N_2(\zeta ) \rangle =c^N_2(1-2\zeta _0)+\bar{c}_2^P 2\zeta _0,
\\
e^L(\zeta _0)&=\langle \frac{1}{\epsilon^P_{33}(\zeta )} \rangle ^{-1} q(1-2\zeta _0).
\end{align*}
Varying functional \eqref{7.0}, we obtain the Euler equation
\begin{equation}
\begin{split}
\bar{\rho }\ddot{u}_\varrho =(\bar{c}^P_1+2\bar{c}^P_2)(u_{\varrho ,\varrho \varrho }+\frac{u_{\varrho ,\varrho }}{\varrho}-\frac{u_\varrho }{\varrho ^2}) \quad \text{for $\varrho
\in (0,r_0)$},
\\
\rho ^L \ddot{u}_\varrho =(c_1^L+2c_2^L)(u_{\varrho ,\varrho \varrho }+\frac{u_{\varrho ,\varrho 
}}{\varrho}-\frac{u_\varrho }{\varrho ^2})\quad \text{for $\varrho
\in (r_0,r)$},
\end{split}
\label{7.1}
\end{equation}
the jump conditions at $\varrho =r_0$
\begin{align}
&u_\varrho |_{r_0-}=u_\varrho |_{r_0+},
\notag \\
&2\zeta _0[(\bar{c}^P_1+2\bar{c}^P_2)u_{\varrho ,\varrho }+\bar{c}^P_1u_\varrho /\varrho ]|_{r_0-}
\label{7.2} \\
&=[(c_1^L+2c_2^L)u_{\varrho ,\varrho }+c_1^Lu_\varrho /\varrho +
e^L\frac{\varphi _0(t)}{h}]|_{r_0+}, \notag
\end{align}
and the traction-free boundary condition at $\varrho =r$
\begin{eqnarray}
[(c_1^L+2c_2^L)u_{\varrho ,\varrho }+c_1^Lu_\varrho /\varrho +
e^L\frac{\varphi _0(t)}{h}]|_{r}=0.
\label{7.3}
\end{eqnarray}
The voltage $\varphi _0(t)$ is assumed to depend harmonically on $t$, $\varphi _0(t)=\hat{\varphi }_0 \cos (\omega t)$, so that solutions of \eqref{7.1}-\eqref{7.3} can be sought in the form 
\begin{equation}
\label{7.4}
u_\varrho =\hat{u}\cos(\omega t).
\end{equation}
Introducing the dimensionless variable and quantities
\begin{gather}
y=\frac{\varrho }{r}, \quad \vartheta =\omega r \sqrt{\frac{\rho ^L}{c^L_1+2c^L_2}}, \quad \lambda =\sqrt{\frac{\rho ^L/\bar{\rho }}{(c^L_1+2c^L_2)/(\bar{c}^P_1+2\bar{c}^P_2)}},
\\
\kappa _1=\frac{2\zeta _0(\bar{c}^P_1+2\bar{c}^P_2)}{c^L_1+2c^L_2}, \quad \kappa _2=\frac{2\zeta _0\bar{c}^P_1}{c^L_1+2c^L_2},\quad \gamma =\frac{c^L_1}{c^L_1+2c^L_2}, \quad d=\frac{e^L}{c^L_1+2c^L_2},
\label{7.5}
\end{gather}
this system can be transformed to the differential equations
\begin{equation*}
\begin{split}
\hat{u}^{\prime \prime }+\frac{1}{y}\hat{u}^\prime 
+(\frac{\vartheta ^2}{\lambda ^2}-\frac{1}{y^2})\hat{u}=0 \quad \text{for $y
\in (0,\bar{r})$},
\\
\hat{u}^{\prime \prime }+\frac{1}{y}\hat{u}^\prime +(\vartheta ^2
-\frac{1}{y^2})\hat{u}=0\quad \text{for $y
\in (\bar{r},1)$},
\end{split}
\end{equation*}
where $\bar{r}=r_0/r$, the jump conditions at $y=\bar{r}$
\begin{gather*}
\hat{u}|_{\bar{r}-}=\hat{u}|_{\bar{r}+},
\\
[\kappa _1\hat{u}^\prime +\kappa _2 \hat{u}/y]|_{\bar{r}-}
=[\hat{u}^\prime +\gamma \hat{u}/y +
d\frac{\hat{\varphi }_0}{h}r]|_{\bar{r}+},
\end{gather*}
and the traction-free boundary condition at $y=1$
\begin{eqnarray*}
[\hat{u}^\prime +\gamma \hat{u}/y +
d\frac{\hat{\varphi }_0}{h}r]|_{y=1}=0.
\end{eqnarray*}

For the part of the plate without piezoceramic patches ($y<\bar{r}$) the solution is given by
\begin{eqnarray*}
\hat{u}(y)=a_1J_1(\frac{\vartheta }{\lambda}y),
\end{eqnarray*}
while for the part covered by the piezoceramic patches ($\bar{r}<y<1$) we have
\begin{eqnarray*}
\hat{u}(y)=a_2J_1(\vartheta y)+a_3Y_1(\vartheta y).
\end{eqnarray*}
Here $J_1(x)$ and $Y_1(x)$ are Bessel function of the first and second kind, respectively.
The constants $a_1,a_2,a_3$ can be determined from the jump conditions at $y=\bar{r}$ and the boundary condition at $y=1$. Substituting the formulas for $\hat{u}(y)$ into them, we get the system of linear equations
\begin{eqnarray*}
\sum_{j=1}^3C_{ij}a_j=b_i,\quad i=1,2,3,
\end{eqnarray*}
where $C_{ij}$ and $b_i$ are given by
\begin{align*}
C_{11}&=J_1(\vartheta \frac{\bar{r}}{\lambda}),\quad 
C_{12}=-J_1(\vartheta \bar{r}),\quad C_{13}=-Y_1(\vartheta \bar{r}),
\\
C_{21}&=(\kappa _2-\kappa _1)J_1(\vartheta \frac{\bar{r}}{\lambda}) 
+\kappa _1\vartheta \frac{\bar{r}}{\lambda}J_0(\vartheta 
\frac{\bar{r}}{\lambda}),
\\
C_{22}&=(1-\gamma )J_1(\vartheta \bar{r})-\vartheta \bar{r}J_0(\vartheta \bar{r}),
\\
C_{23}&=(1-\gamma )Y_1(\vartheta \bar{r})-\vartheta \bar{r}Y_0(\vartheta \bar{r}),
\\
C_{31}&=0,\quad C_{32}=(1-\gamma )J_1(\vartheta )-\vartheta J_0(\vartheta ),
\\
C_{33}&=(1-\gamma )Y_1(\vartheta )-\vartheta Y_0(\vartheta ),
\\
b_1&=0,\quad b_2=d\frac{\hat{\varphi }_0}{h}r\bar{r},
\quad b_3=d\frac{\hat{\varphi }_0}{h}r.
\end{align*}

After finding $a_i$ we can determine the amplitude of $D^3$ by the formula
\begin{align*}
\hat{D}^3&=-\langle \frac{1}{\epsilon^P_{33}(\zeta )}\rangle ^{-1}\frac{\hat{\varphi }_0}{h}
+\langle \frac{1}{\epsilon^P_{33}(\zeta )}\rangle ^{-1}q(1-2\zeta _0)(\hat{u}_{,\varrho }+\frac{\hat{u}}{\varrho})
\\
&=-\langle \frac{1}{\epsilon^P_{33}(\zeta )}\rangle ^{-1}\frac{\hat{\varphi }_0}{h}
+\langle \frac{1}{\epsilon^P_{33}(\zeta )}\rangle ^{-1}q(1-2\zeta _0)[a_2\eta J_0(\eta \varrho 
)+a_3\eta Y_0(\eta \varrho )],
\end{align*}
where $\eta =\vartheta /r$. Then the amplitude of the total charge on one of the electrodes is equal to
\begin{align*}
\int_{\Omega_1}\hat{D}^3\, da&=-\pi r^2\langle \frac{1}{\epsilon^P_{33}(\zeta )}\rangle ^{-1}\frac{\hat{\varphi }_0}{h}\{ (1-\bar{r}^2)-
2qd(1-2\zeta _0)[\bar{a}_2(J_1(\vartheta )-\bar{r}J_1(\vartheta \bar{r}))
\\
&+\bar{a}_3(Y_1(\vartheta )-\bar{r}Y_1(\vartheta \bar{r}))]\} ,
\end{align*}
where $\bar{a}_i$ is the solution of the system
\begin{eqnarray*}
\sum_{j=1}^3C_{ij}\bar{a}_j=\bar{b}_i,\quad i=1,2,3,
\end{eqnarray*}
with 
\[
\bar{b}_1=0,\quad \bar{b}_2=\bar{r},\quad \bar{b}_3=1.
\]

According to the solution of this problem the resonant frequencies are the roots of the determinantal equation
\begin{equation}
\det C_{ij} =0.
\label{7.6}
\end{equation}
The antiresonant frequencies should be found from the condition that the total charge vanishes giving
\begin{equation}
\frac{1-\bar{r}^2}{2qd(1-2\zeta _0)}=\bar{a}_2[J_1(\vartheta 
)-\bar{r}J_1(\vartheta \bar{r})]
+\bar{a}_3[Y_1(\vartheta )-\bar{r}Y_1(\vartheta \bar{r})].
\label{7.7}
\end{equation}

\begin{table}[htb]
  \centering
\begin{tabular}{|l|c|c|c|c|c|c|c|c|c|c|c|} \hline
 $s_{11}^E$ & $s^E_{13}$ &
  $s^E_{33}$ & $s^E_{55}$ & $s^E_{66}$ & $d_{15}$ & $d_{31}$ & $d_{33}$ & $\epsilon ^T_{11}/\epsilon _0$ & $\epsilon ^T_{33}/\epsilon _0$ & $\rho $ \\ \hline
16.4 & -7.22 & 18.8 & 47.5 & 44.3 & 584 & -171 & -374 & 1730 & 1700 & 7.75 \\ \hline
\end{tabular}
  \caption{Material constants of PZT-5A}
  \label{table1}
\end{table}

For the numerical simulations we use as an example aluminum oxide (Al$_2$O$_3$) as the isotropic elastic (dielectric and non-conducting) material and PZT-5A polarized in the third direction as the piezoceramic material. The material constants for PZT-5A, taken from \citep{Berlincourt1964}, are presented in Table.~\ref{table1}, with $\epsilon _0=8.854 \times 10^{-12}$F/m being the dielectric constant of vacuum. In this Table the dimension unit for $s^E_{\mathfrak{m} \mathfrak{n}}$ is $10^{-12}$m$^2$/N, for $d_{a \mathfrak{m}}$ is $10^{-12}$C/N, and for $\rho $ is $10^{3}$kg/m$^3$. Besides, for piezoceramic material we have \citep{Le1999}
\begin{equation}
\label{7.9}
\begin{split}
c^N_1=\frac{1}{2s^E_{11}(1+\nu )}\frac{\nu +\frac{1-\nu }{2}k_p^2}{1-k_p^2}, \quad  c_2^N=\frac{1}{2s^E_{11}(1+\nu )}, \quad \varepsilon ^P_{33}=\epsilon ^T_{33}(1-k_p^2)
\\
q=\frac{e^{31}_P}{\epsilon^{33}_P}=\frac{1}{\epsilon^S_{33}(1+k_t^2)}\frac{d_{31}}{s^E_{11}(1-\nu )}, \quad k_p^2=\frac{2d^2_{31}}{(1-\nu )\epsilon ^T_{33}s^E_{11}},
\end{split}
\end{equation}
where $\nu =-(s^E_{11}-s^E_{66}/2)/s^E_{11}$ is Poisson's ratio and $k_p^2$ the planar coupling factor. From Table.~\ref{table1} we find that $\nu =0.35$ and $k_p=0.6$.

\begin{figure}[htb]
    \begin{center}
    \includegraphics[height=7cm]{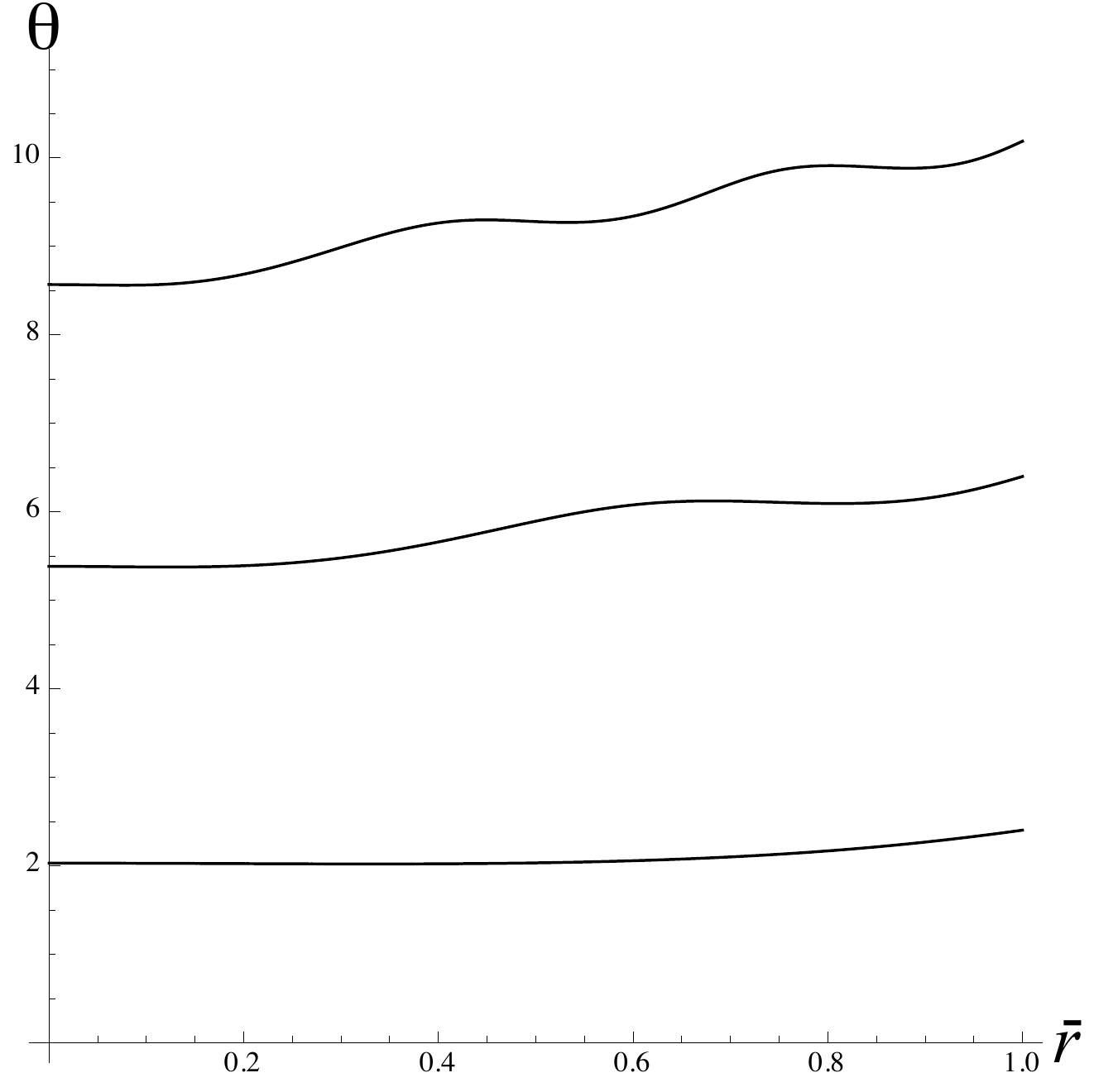}
    \end{center}
    \caption{Three first resonant frequencies of the circular plate partially covered by piezoceramic rings versus the relative radius $\bar{r}$ of the uncovered area at fixed $\zeta _0=0.4$.}
    \label{fig:2}
\end{figure}

The material constants of aluminum oxide, taken from \citep{Touloukian1966}, are
\begin{equation}
\label{7.8}
\bar{E}=385\text{GPa}, \quad \bar{\nu }=0.254, \quad \bar{\rho }=3.9\times 10^{3}\text{kg/m}^3, \quad \bar{\epsilon }_{33}/\epsilon _0=11.54.
\end{equation}
Furthermore, for an elastic (dielectric) isotropic material
\begin{equation}
\label{7.10}
\bar{c}^P_1=2\bar{\mu }\bar{\sigma }=\frac{\bar{E}\bar{\nu }}{1-\bar{\nu }^2}, \quad \bar{c}^P_2=\bar{\mu }=\frac{\bar{E}}{2(1+\bar{\nu })},\quad \bar{\epsilon }^P_{33}=\bar{\epsilon }_{33}.
\end{equation}

\begin{figure}[htb]
    \begin{center}
    \includegraphics[height=7cm]{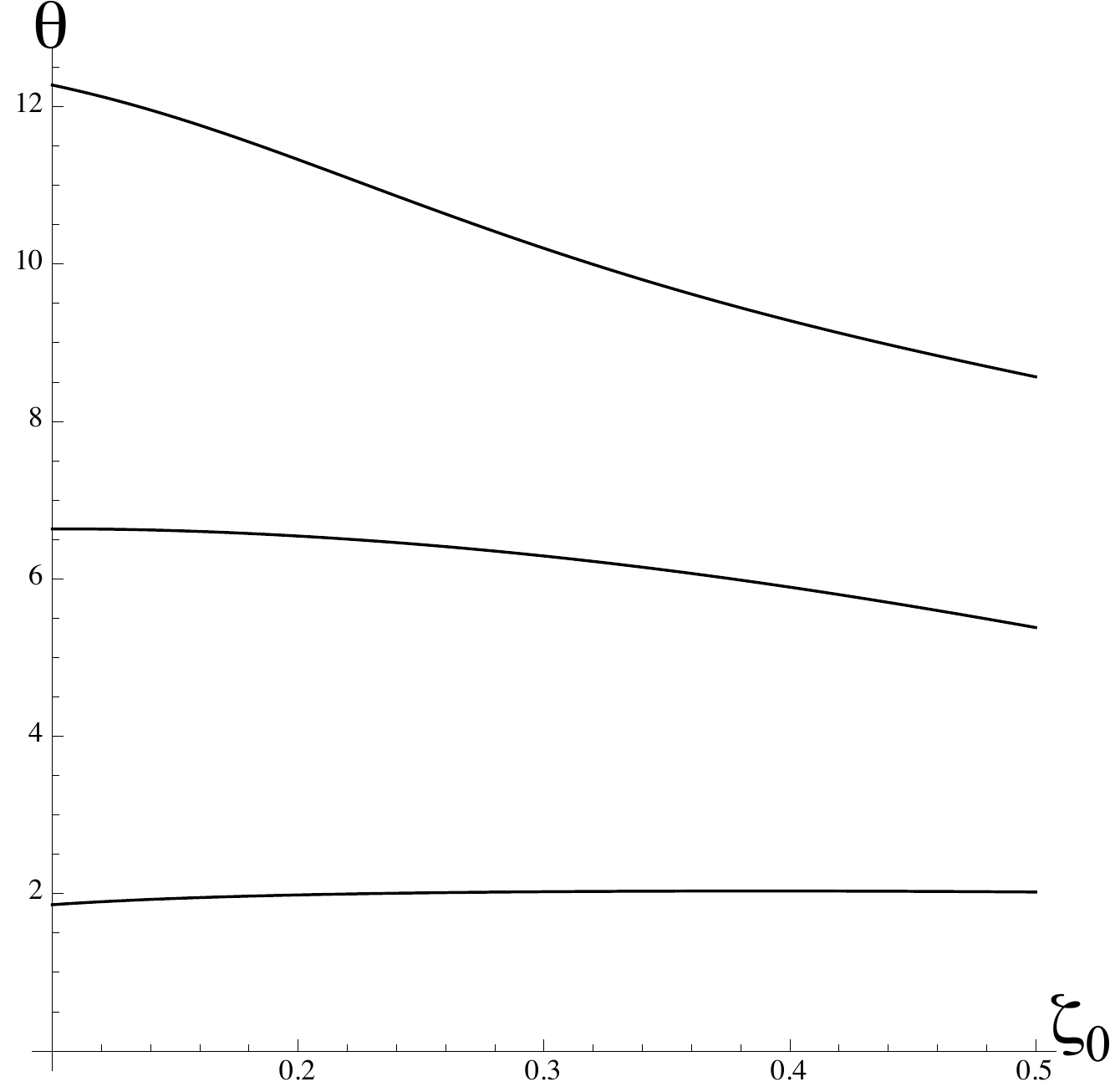}
    \end{center}
    \caption{Three first resonant frequencies of the circular plate partially covered by piezoceramic rings versus the relative thickness $\zeta _0$ at fixed $\bar{r}=0.5$.}
    \label{fig:3}
\end{figure}
 
With these material data we can compute the numerical values of constants used in this 2-D problem. The results of numerical simulations are shown in Figures 2, 3, and 4. Figure 2 plots the first three roots of equations \eqref{7.6} as functions of the parameter $\bar{r}=r_0/r$ for $\zeta _0=0.4$ representing the resonant frequencies of the smart sandwich plate versus $\bar{r}$. Figure 3 shows the first three roots of equations \eqref{7.6} as functions of the parameter $\zeta _0$ changing between 0.1 and 0.5 at fixed $\bar{r}=0.5$ representing the resonant frequencies versus $\zeta _0$. It is interesting to note that, in the limit $\bar{r}\to 0$ and $\zeta _0\to 0$, the resonant frequencies tend to those of the piezoceramic homogeneous plate covered by the electrodes, while in the limit $\bar{r}\to 1$ and $\zeta _0\to 1/2$ they tend to those of the purely elastic plate.

\begin{figure}[htb]
    \begin{center}
    \includegraphics[height=7cm]{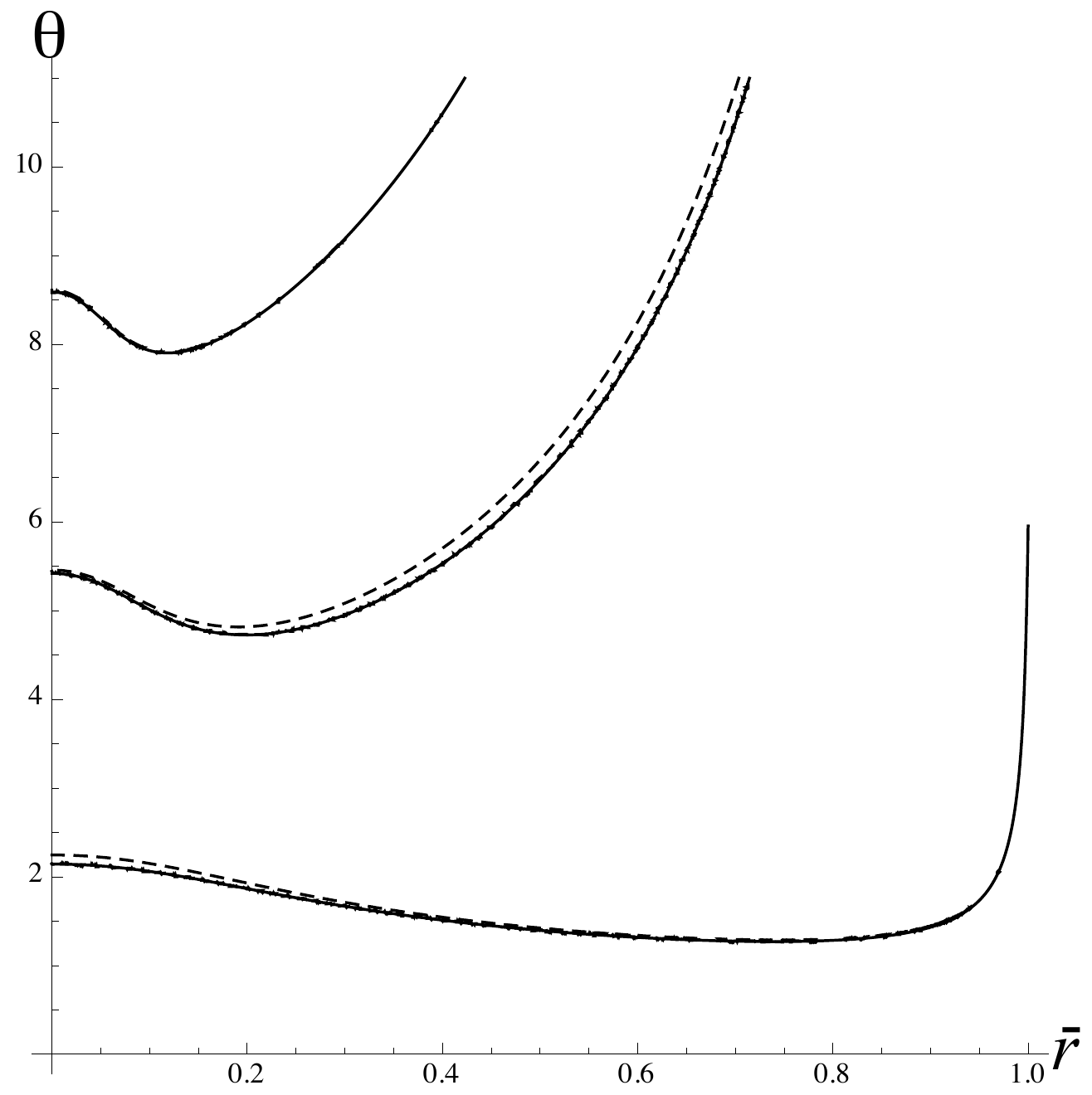}
    \end{center}
    \caption{Three first resonant (bold lines) and anti-resonant frequencies (dashed lines) of the circular plate partially covered by piezoceramic rings versus $\bar{r}$ at fixed $\zeta _0=0.01$.}
    \label{fig:4}
\end{figure}

Concerning the anti-resonant frequencies we observe that they lye slightly above the resonant ones only if $\zeta _0$ is close to zero as show in Fig.~4 for $\zeta _0=0.01$, where the resonant frequencies correspond to the bold lines, while the anti-resonant frequencies to the dashed line. This is close to the behavior of the purely piezoceramic plate \cite{Le1999}. When $\zeta _0$ is close to 1/2 which means that the piezoceramic patches are thin compared to the thickness of the elastic layer, the anti-resonant frequencies are much higher than the resonant ones. Thus, the sandwich plate with thin piezoceramic patches does not have electrically anti-resonant vibrations near the resonant frequencies.

\section{Conclusion}
It is shown in this paper that the rigorous first order approximate 2-D theory of thin smart  sandwich shells can be derived from the exact 3-D piezoelectricity theory by the variational-asymptotic method. The strains and electric field of the smart sandwich shell turn out to be  discontinuous through the thickness and differ essentially from those of the homogeneous piezoelectric shells. The error estimation for the constructed 2-D theory is established that enables one to apply this theory to the problem of vibration control of thin elastic shells with bonded piezoelectric patches. Note that the elastic layer of smart structures used in engineering praxis (like epoxy/glass material) may have more complicated anisotropy property for which the theory developed here is not applicable. Our study in such cases is still going on and the results will be reported elsewhere.

\end{document}